\documentclass[twocolumn,showpacs,preprintnumbers,amsmath,amssymb,aps,prb,superscriptaddress]{revtex4-2}

\pdfoutput=1
\usepackage[pdftex]{graphicx}
\usepackage[english]{babel}
\usepackage{soul}
\usepackage{cleveref}
\usepackage{bbold}
\usepackage{mathtools}
\usepackage{multirow}
\usepackage{colortbl}
\definecolor{kugray5}{RGB}{224,224,224}
\usepackage[latin1]{inputenc}
\usepackage{diagbox}
\usepackage{mciteplus}

\usepackage{dcolumn}
\usepackage{bm}

\usepackage{color}
\usepackage{amstext}
\usepackage{braket}
\usepackage{mathdots}
\usepackage{tikz}
\usepackage{physics}
\usepackage{enumitem}

\usetikzlibrary{matrix,decorations.pathreplacing}

\begin{document}


\title{Kaleidoscopes of Hofstadter Butterflies and Aharonov-Bohm caging from  $2^n$-root topology in decorated square lattices}

\author{A. M. Marques}
\email{anselmomagalhaes@ua.pt}
\affiliation{Department of Physics $\&$ i3N, University of Aveiro, 3810-193 Aveiro, Portugal}

\author{J. M\"ogerle}
\affiliation{Institute for Theoretical Physics III and Center for Integrated Quantum Science and Technology, University of Stuttgart, 70550 Stuttgart, Germany}
\affiliation{Department of Physics and SUPA, University of Strathclyde, Glasgow G4 0NG, United Kingdom}

\author{G. Pelegr\'{\i}}
\affiliation{Department of Physics and SUPA, University of Strathclyde, Glasgow G4 0NG, United Kingdom}

\author{S. Flannigan}
\affiliation{Department of Physics and SUPA, University of Strathclyde, Glasgow G4 0NG, United Kingdom}

\author{R. G. Dias}
\affiliation{Department of Physics $\&$ i3N, University of Aveiro, 3810-193 Aveiro, Portugal}

\author{A. J. Daley}
\affiliation{Department of Physics and SUPA, University of Strathclyde, Glasgow G4 0NG, United Kingdom}


\begin{abstract}
Square-root topology describes models whose topological properties can be revealed upon squaring the Hamiltonian, which produces their respective parent topological insulators. 
This concept has recently been generalized to $2^n$-root topology, characterizing models where $n$ squaring operations must be applied to the Hamiltonian in order to arrive at the topological source of the model.
In this paper, we analyze the 
Hofstadter regime of quasi-one-dimensional (quasi-1D) and two-dimensional (2D) $2^n$-root models, the latter of which has the square lattice (SL) (known for the Hofstadter Butterfly) as the source model.
We show that upon increasing the root-degree of the model, there appear multiple magnetic flux insensitive flat bands, and analytically determine corresponding eigenstates. These can be recast as compact localized states (CLSs) occupying a finite region of the lattice. For a finite flux, these CLSs correspond to different harmonics contained within the same Aharonov-Bohm (AB) cage.
Furthermore, as the root-degree increases, a kaleidoscope of butterflies is seen to appear in the Hofstadter diagram, with each butterfly constituting a topologically equivalent replica of the original one of the SL.
As such, the index $n$, which uniquely identifies the root-degree of the model, can be seen as an additional fractal dimension of the $2^n$-root model present in its Hofstadter diagram.
We discuss how these dynamics could be realized in experiments with ultracold atoms, and measured by Bragg spectroscopy or through observing dynamics of initially localized atoms in a quantum gas microscope. 

\end{abstract}

\pacs{74.25.Dw,74.25.Bt}

\maketitle
\section{Introduction}
\label{sec:intro}

\begin{figure}[tb]
	\begin{centering}
		\includegraphics[width=0.48 \textwidth]{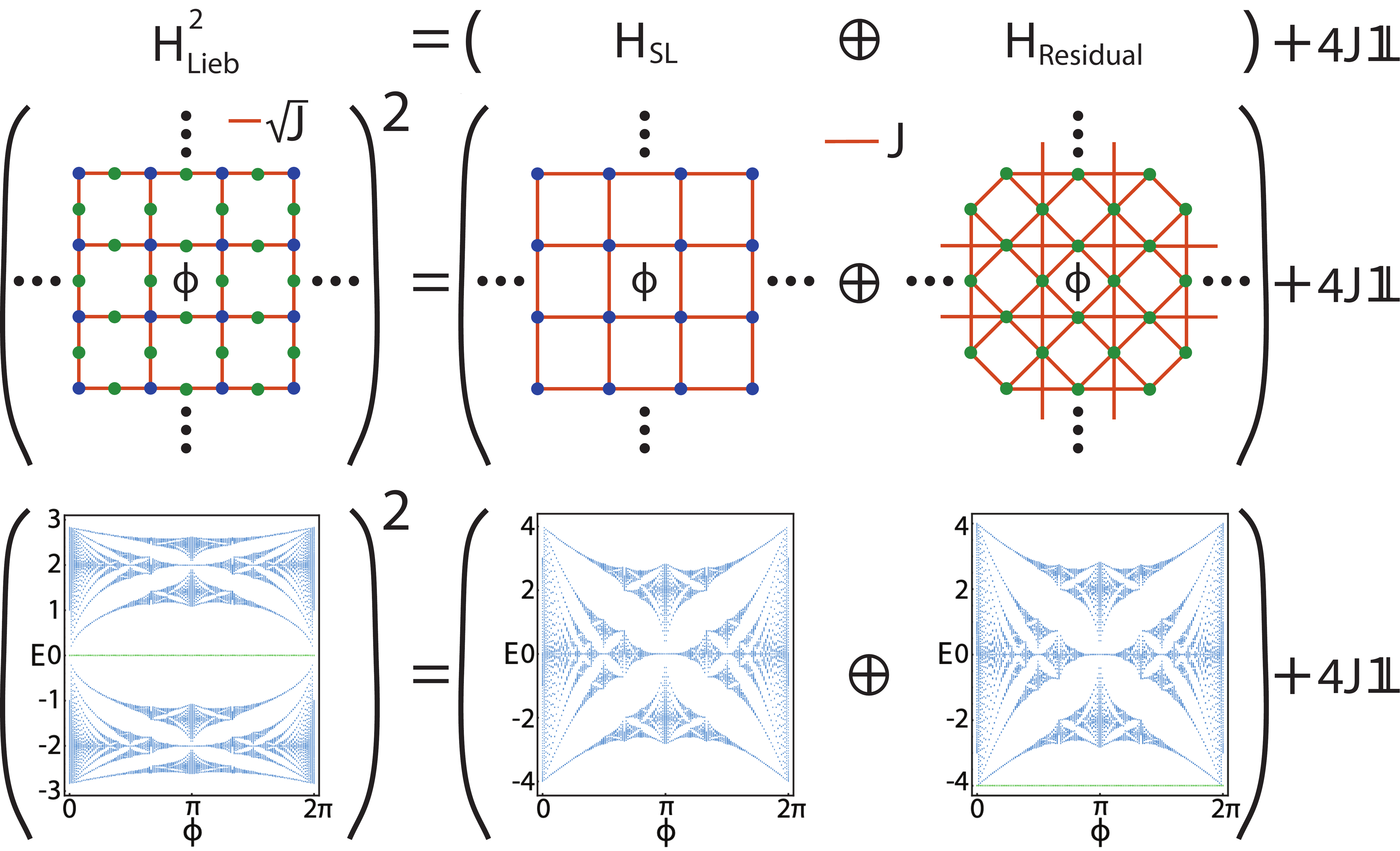}
		\par\end{centering}
	\caption{Illustration of the squaring process applied to the Lieb lattice with a $\phi$ magnetic flux per plaquette. On the level of the Hamiltonian (top) squaring leads to two diagonal blocks, corresponding to the Hamiltonians of two decoupled lattices. This corresponds (middle) to the SL on the sublattice shown in blue on the original Lieb lattice, and a residual lattice on the remaining green sublattice sites, with both keeping the $\phi$ flux per plaquette, as well as a global energy shift of $4J$. Below this, we show the corresponding Hofstadter spectra with periodic boundary conditions, with $J\equiv 1$ as the energy unit and the dispersive (flat) bands are colored in blue (green).}
	\label{fig:1}
\end{figure}

Topological insulators (TIs) \cite{Hasan2010,Araujo2021}, describing models with insulating bulk properties and topologically protected metallic surface states, have been one of the most active research topics of the past two decades in condensed matter physics.
An interesting recent development has been that of square-root TIs ($\sqrt{\text{TIs}}$) \cite{Arkinstall2017,Kremer2020,Pelegri2019,Mizoguchi2020,Ezawa2020,Ke2020,Lin2021,Yoshida2021,Wu2021,Ding2021,Geng2021,Zhang2022,Matsumoto2022} and square-root topological semimetals \cite{Mizoguchi2021c,Palmer2022}.
These systems require the application of non-linear algebraic operations in order for them to manifest their topological nature.
In simpler terms, one has to square the Hamiltonian of a $\sqrt{\text{TI}}$ to find the original TI as one of its diagonal blocks, from which the $\sqrt{\text{TI}}$ inherits its topological properties, leading to new phenomena such as fractional topological weight on their finite energy states.
This process is illustrated in Fig.~\ref{fig:1} where, anticipating part of our results below, we schematically demonstrate how the Lieb lattice is the square-root model of the SL, and how the squaring operation remarkably relates the Hofstadter spectrum of both models.

Several experimental realizations of these models are already available, whether on acoustic \cite{Yan2020,Cheng2022}, electrical \cite{Song2020,Song2022} or photonic \cite{Kremer2020,Yan2021,Kang2021} lattices.
The intriguing possibility of addressing supersymmetry in a condensed matter context through square-root topology has also started to be explored \cite{Roy2022}.
At the same time, related work on multiplicative topological phases which, analogously to $\sqrt{\text{TIs}}$, require two parent systems from which a child topological insulator is derived following a novel procedure, is already underway \cite{Cook2022}.

Other recent studies have already discussed higher-root generalizations of square-root topology \cite{Dias2021,Marques2021,Marques2021b,Deng2022}, including in Floquet systems \cite{Bomantara2022,Zhou2022} and for fractional powers of Bogoliubov-de Gennes Hamiltonians \cite{Basa2022}.
In particular, $2^n$-root TIs ($\sqrt[2^n]{\text{TIs}}$) \cite{Dias2021,Marques2021,Marques2021b}, with $n\in\mathbb{N}$, are a direct generalization of $\sqrt{\text{TIs}}$, since for these models $n$ successive squaring operations have to be performed on the Hamiltonian in order to arrive at the original TI.
A proliferation of edge states at different energy gaps was seen to occur for these $\sqrt[2^n]{\text{TIs}}$, each of which exhibiting a diluting overlap with the topological state of the TI with increasing $n$.

However, up to now the effects of introducing magnetic flux on these high-root topological systems have not been addressed.
In this work, we will fill this gap by addressing the Hofstadter diagram of both a quasi-1D $\sqrt[2^n]{\text{TIs}}$, whose original TI is the Creutz ladder (CL) \cite{Marques2021}, and a 2D lattice, whose original TI is the SL, well-known for its Hofstadter butterfly \cite{Hofstadter1976,Naumis2016}, recently observed experimentally in photonic lattices \cite{Weidemann2022} and electrical circuits \cite{Yatsugi2022}.
We will provide three main results related to the Hofstadter diagram of these $\sqrt[2^n]{\text{TIs}}$: (i) the existence of multiple flux insensitive flat bands, all inert with respect to their strong topological indices; (ii) an analytical derivation of all flat band eigenstates, with or without flux, also in the form of CLSs, which are shown to share a common AB cage when flux is applied; and (iii)
the replication of the butterfly of the SL, with the exact same topological properties, in the Hofstadter diagram as $n$ increases, which introduces a \textit{second} fractal dimension to the problem, which also determines the number of butterflies in the kaleidoscope.

We point out that a different scheme for replicating the Hofstadter butterfly has also been proposed for monolayer \cite{Arora2018} and bilayer \cite{Arora2022} graphene under periodic modulations of the magnetic field.
Our work is further stimulated by the renewed interest shown as of lately in the Hofstadter regime of different models, including dimerized topological lattices \cite{Lau2015,Zuo2021}, models with fractal defects \cite{Matsuki2021}, Weyl semimetals \cite{Abdulla2022}, twisted bilayer and trilayer graphene \cite{Lian2020,Herzog2020,Lu2021,Sheffer2021,Imran2022}, non-Hermitian systems \cite{Shao2022}, or quasicrystals \cite{Ghadimi2022}.

Beyond the experimental platforms mentioned before, neutral atoms manipulated with light fields are an excellent candidate to implement the $2^n$-root models that we discuss in this work.
Ultracold atoms in optical lattices have already been used to realize the Hofstadter Hamiltonian (the SL with a finite synthetic magnetic flux per plaquette) \cite{Aidelsburger2013,Miyake2013,Aidelsburger2015}, and the tools that were used in these settings to implement fluxes, such as Raman-assisted tunneling processes \cite{PhysRevLett.107.255301}, could be applied to the more sophisticated geometries required for $2^n$-root topology.
Neutral atoms manipulated with optical tweezers and excited to Rydberg states  have also been used to study topological models \cite{de2019observation,weber2018topologically} and allow for the realization of synthetic fluxes \cite{lienhard2020realization}.
With these exciting perspectives in mind, we propose a protocol based on Bragg spectroscopy of non-interacting atoms in optical lattices to probe the repeating band structure that arises in 2D $2^n$-root models.
Furthermore, we show how the Fourier analysis of the dynamics of single atoms prepared in specific lattice sites can reveal information about the multiple flat bands and the AB caging effect.

The rest of the paper is organized as follows.
In Sec.~\ref{sec:2ncl}, we motivate the main results appearing later on by addressing first the topological properties, the analytical derivation of the flat band subspace and the Hofstadter diagram of a $2^n$-root family of quasi-1D models.
In Sec.~\ref{sec:2nrootsl}, we introduce a 2D family of $2^n$-root models, whose parent model is given by the SL, and study the fluxless regime, similarly providing an exact analytic treatment of the flat band subspace.
In Sec.~\ref{sec:hofst}, the Hofstadter spectrum of these 2D models is provided, and the Chern topology at each energy gap is determined.
Still in this section, we derive the general form of the CLSs of the different flat bands for an arbitrary flux, which are shown to originate from an AB caging mechanism.
In Sec.~\ref{sec:exp_impl}, we describe experimental schemes that can be implemented to both realize these $2^n$-root models and probe their distinctive features, focusing in particular on optical lattices due to their high versatility.
Finally, we present our conclusions in Sec.~\ref{sec:conclusions}.

\section{Decorated diamond chain with flux}
\label{sec:2ncl}
We begin by considering a quasi-1D model, namely the $2^n$-root version of the CL ($\sqrt[2^n]{\text{CL}}$), depicted in Fig.~\ref{fig:2}, first introduced in the Appendix~A of \cite{Marques2021}.
It will be convenient for our discussion below to work with two different bases for the Hilbert space,
\begin{eqnarray}
	\mathcal{B}_A&:=& \ \ \{\ket{j,k}\},\ \ \  j=1,2,\dots,2^{n+1}-1,
	\label{eq:basisa}
	\\
	\mathcal{B}_B&:=&\{\{\text{BS}\},\{\text{GS}\}\},
	\label{eq:basisb}
\end{eqnarray}
where $\mathcal{B}_A$ is the site ordered basis (shown in the zoomed unit cell in Fig.~\ref{fig:2}) and $\mathcal{B}_B$ is the manifestly chiral-symmetric basis, with BS (GS) labeling the ordered sites within the blue (green) sublattice of the unit cell.
Under periodic boundary conditions (PBC), the bulk Hamiltonian for each $n$\textit{}, written in the $\mathcal{B}_A$ basis and with lattice constant set to $a\equiv1$, is given by
	\begin{equation}
		H_{\sqrt[2^n]{\text{CL}},A}(k)=
		\begin{pmatrix}
			0&\mathbf{v}^0_{2^n}(k)&\mathbf{v}^\phi_{2^n}(k)
			\\
			\mathbf{v}^{0\dagger}_{2^n}(k)&H^{\text{CL}}_{\text{Linkage},2^n}&O_{2^{n}-1}
			\\
			\mathbf{v}^{\phi\dagger}_{2^n}(k)&O_{2^{n}-1}&H^{\text{CL}}_{\text{Linkage},2^n}&
		\end{pmatrix},
	\label{eq:hamilt2ncl}
	\end{equation}
where the subscript $\mu=A,B$ indicates the basis $\mathcal{B}_\mu$ used, $\mathbf{v}^\phi_{2^n}(k)=-\sqrt[2^n]{J}(1,0,\dots,0,e^{-i(k+\phi)})$ is a vector of size $2^{n}-1$, $O_j$ is the zero square-matrix of size $j$, and $H^{\text{CL}}_{\text{Linkage},2^n}$ is the tridiagonal Hamiltonian of an open linear chain of $2^n-1$ sites and uniform nearest-neighbor couplings $\sqrt{2}\sqrt[2^n]{J}$.
Note that the entire Peierls phase is accumulated in the last element of the first row, that is, in the hopping between the last site $2^{n+1}-1$ and the spinal site 1.
We set $J$ as the energy unit for the rest of the paper and the reduced Planck's constant to $\hbar\equiv1$.
As an example, the energy spectrum for $n=3$, obtained from diagonalization of (\ref{eq:hamilt2ncl}), is shown in Fig.~\ref{fig:3}(a), where eight dispersive bands are seen to intercalate with seven flat bands, with triple band touching points occurring at the inversion-invariant momenta $k=0,\pi$ in alternation.

In Fig.~\ref{fig:3}(b), we show the energy spectrum of the same periodic $\sqrt[8]{\text{CL}}$, but with a finite size of $N_{uc}=30$ unit cells and as a function now of the flux per plaquette.
Its most prominent feature is the persistence of the seven flux insensitive flat bands, which are gapped for $\phi\neq 0$ and whose energies coincide with the eigenvalues of the linear chain described by $H^{\text{CL}}_{\text{Linkage},8}$ in (\ref{eq:hamilt2ncl}),
\begin{equation}
	\epsilon_j=-2\sqrt{2}\cos(\frac{\pi}{2^{n}}j),
	\label{eq:eigvals}
\end{equation}
with $ j=1,2,\dots,2^{n}-1$.
This behavior strongly indicates that these flat bands form a subspace whose eigenstates span over the linkages, which include the shared green sites with the linker, but have nodes at the spinal site of the linker (see Fig.~\ref{fig:2}).
Following [\onlinecite{Mizoguchi2021b}], the proof of the shared eigenvalues between $H^{\text{CL}}_{\text{Linkage},2^n}$ and the flat band subspace of $H_{\sqrt[2^n]{\text{CL}}}(k)$ relies on the existence of a nontrivial rectangular matrix $G_{2^n}^\phi(k)$ of size $(2^{n+1}-1)\times( 2^{n}-1)$, called the intertwiner \cite{Difrancesco1990,Pearce1993}, such that the following identity holds,
\begin{equation}
	H_{\sqrt[2^n]{\text{CL}}}(k)G_{2^n}^\phi(k)=G_{2^n}^\phi(k)H^{\text{CL}}_{\text{Linkage},2^n},
	\label{eq:intertwiner}
\end{equation}
and the matrix resulting from the multiplication is nontrivial.
\begin{figure}[ht]
	\begin{centering}
		\includegraphics[width=0.48 \textwidth]{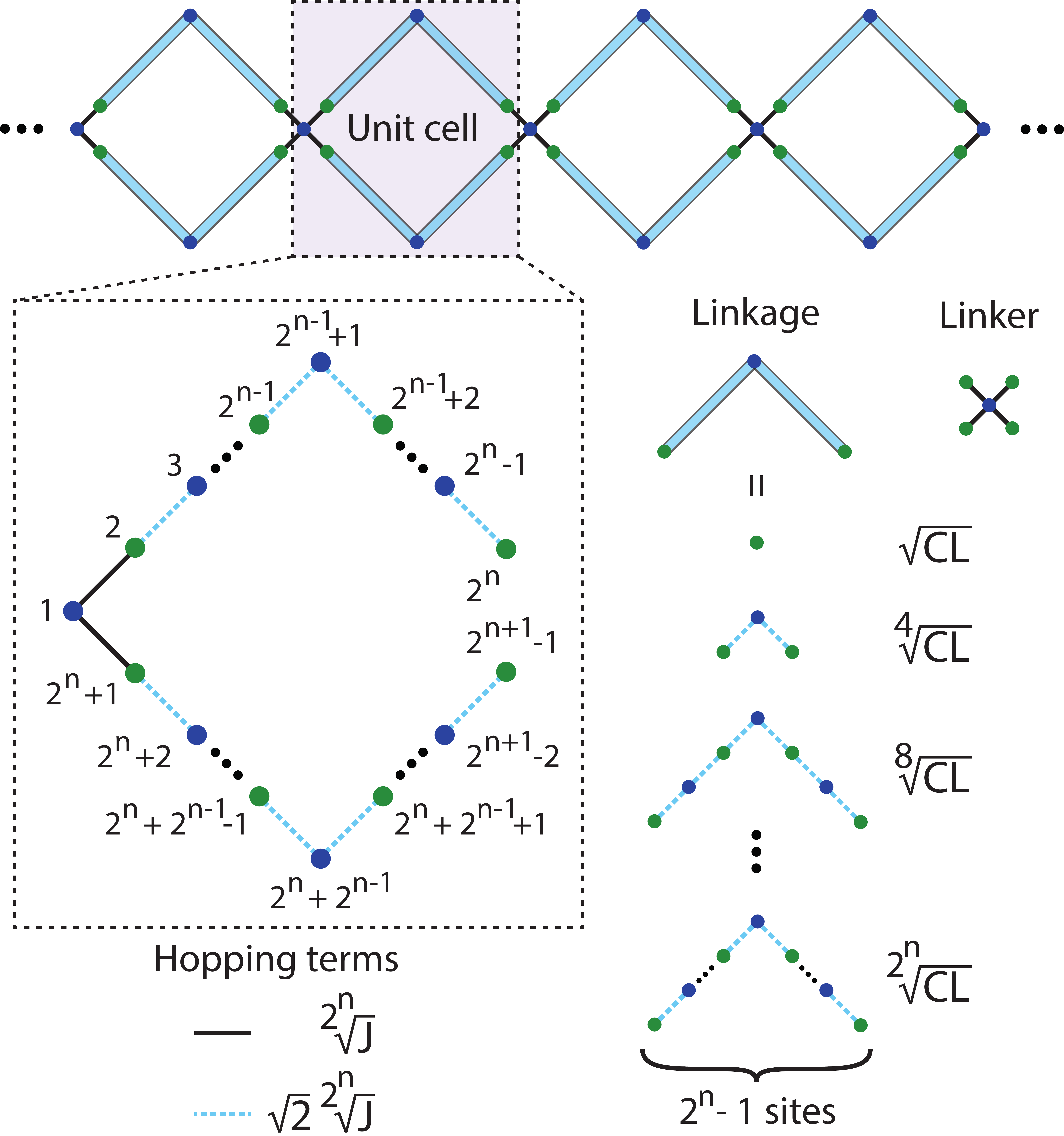}
		\par\end{centering}
	\caption{Illustration of the $\sqrt[2^n]{\text{CL}}$.
		The four green sites in the linker are shared with the adjacent linkages.
		The blue sublattice for a given $n$ constitutes both sublattices of the $n-1$ chain.
		}
	\label{fig:2}
\end{figure}
It readily follows from this identity that the eigenvalues of the linkage given in (\ref{eq:eigvals}), whose correspondent eigenstates satisfy
\begin{equation}
	H^{\text{CL}}_{\text{Linkage},2^n}\varphi_{2^n,j}=\epsilon_j\varphi_{2^n,j},
	\label{eq:schroedlinkagecl}
\end{equation}
are also eigenvalues of $H_{\sqrt[2^n]{\text{CL}}}(k)$, since applying $G_{2^n}^\phi(k)$ on both sides of (\ref{eq:schroedlinkagecl}) from the left leads to
\begin{eqnarray}
	G_{2^n}^\phi(k)H^{\text{CL}}_{\text{Linkage},2^n}\varphi_{2^n,j}&=&\epsilon_jG_{2^n}^\phi(k)\varphi_{2^n,j} \nonumber
	\\
	H_{\sqrt[2^n]{\text{CL}}}(k)\big(G_{2^n}^\phi(k)\varphi_{2^n,j}\big)&=&\epsilon_j\big(G_{2^n}^\phi(k)\varphi_{2^n,j}\big) \nonumber
	\\
	H_{\sqrt[2^n]{\text{CL}}}(k)\psi_{2^n,j}(k)&=&\epsilon_j\psi_{2^n,j}(k),
	\label{eq:commoneigs}
\end{eqnarray}
where (\ref{eq:intertwiner}) was used in the second line, $\varphi_{2^n,j}\not\in\ker G_{2^n}^\phi(k)$ is assumed and
\begin{equation}
	\psi_{2^n,j}(k)=\frac{1}{\mathcal{N}_{2^n}^j(k)}G_{2^n}^\phi(k)\varphi_{2^n,j}
	\label{eq:bulkflateigs}
\end{equation}
 is the $j^\text{th}$ eigenstate within the flat band subspace (since $\epsilon_j$ is $k$-independent) of the bulk Hamiltonian $H_{\sqrt[2^n]{\text{CL}}}(k)$, with $\mathcal{N}_{2^n}^j(k)$ the respective normalization constant.
 An intertwiner that satisfies all these conditions can be found to read as
 \begin{equation}
 	G_{2^n}^\phi(k)=\begin{pmatrix}
 		\vec{O}_{2^{n}-1}
 		\\
 		\alpha I_{2^{n}-1}+\gamma J_{2^{n}-1}
 		\\
 		\beta I_{2^{n}-1}+\delta J_{2^{n}-1}
 		\label{eq:intertwinercl}
 	\end{pmatrix},
 \end{equation}
where $\vec{O}_j$, $I_j$ and $J_j$ are the zero vector, the identity and the exchange matrix of size $j$, respectively, and the coefficients are given by
  \begin{eqnarray}
 	\gamma&=&\delta:=e^{-ik}-e^{-i(k+\phi)},
 	\\
 	\alpha&=&e^{-i(2k+\phi)}(1+e^{-i\phi})-2,
 	\\
 	\beta&=&2-e^{-2ik}(1+e^{-i\phi}).
 \end{eqnarray}
In particular, when $\phi=0\ (\text{mod} \ 2\pi)$ these coefficients simplify to $\gamma=\delta=0$ and $\alpha=-\beta=2(e^{-2ik}-1)$, which is an overall constant that can be factored out (and in fact it is convenient to do so, otherwise the intertwiner is ill-defined for $k=0$), such that one arrives at
 \begin{equation}
	G_{2^n}^0=\begin{pmatrix}
		\vec{O}_{2^{n+1}-1}
		\\
		I_{2^{n+1}-1}
		\\
		-I_{2^{n+1}-1}
	\end{pmatrix},
\end{equation}
which is $k$-independent, reflecting the fact that the eigenstates of each band $j$ in (\ref{eq:bulkflateigs}) can be linearly combined, for a periodic chain with $N_{uc}$ unit cells, to form compact localized states (CLSs) occupying a single unit cell \cite{Nicolau2022a}, whereas the CLSs for a finite flux per plaquette spread over two adjacent unit cells \cite{Lopes2014,Pelegri2019,Pelegri2019b}, as a consequence of the irremovable $k$-dependence in $G_{2^n}^\phi(k)$ for $\phi\neq 0$.
However, regardless of the flux value, the zero vector at the first row of the intertwiner in (\ref{eq:intertwinercl}) shows that all flat band states in this subspace have a node on the component at the spinal site, which acts as a dark site \cite{Morales2016} with zero weight.
\begin{figure}[ht]
	\begin{centering}
		\includegraphics[width=0.48 \textwidth]{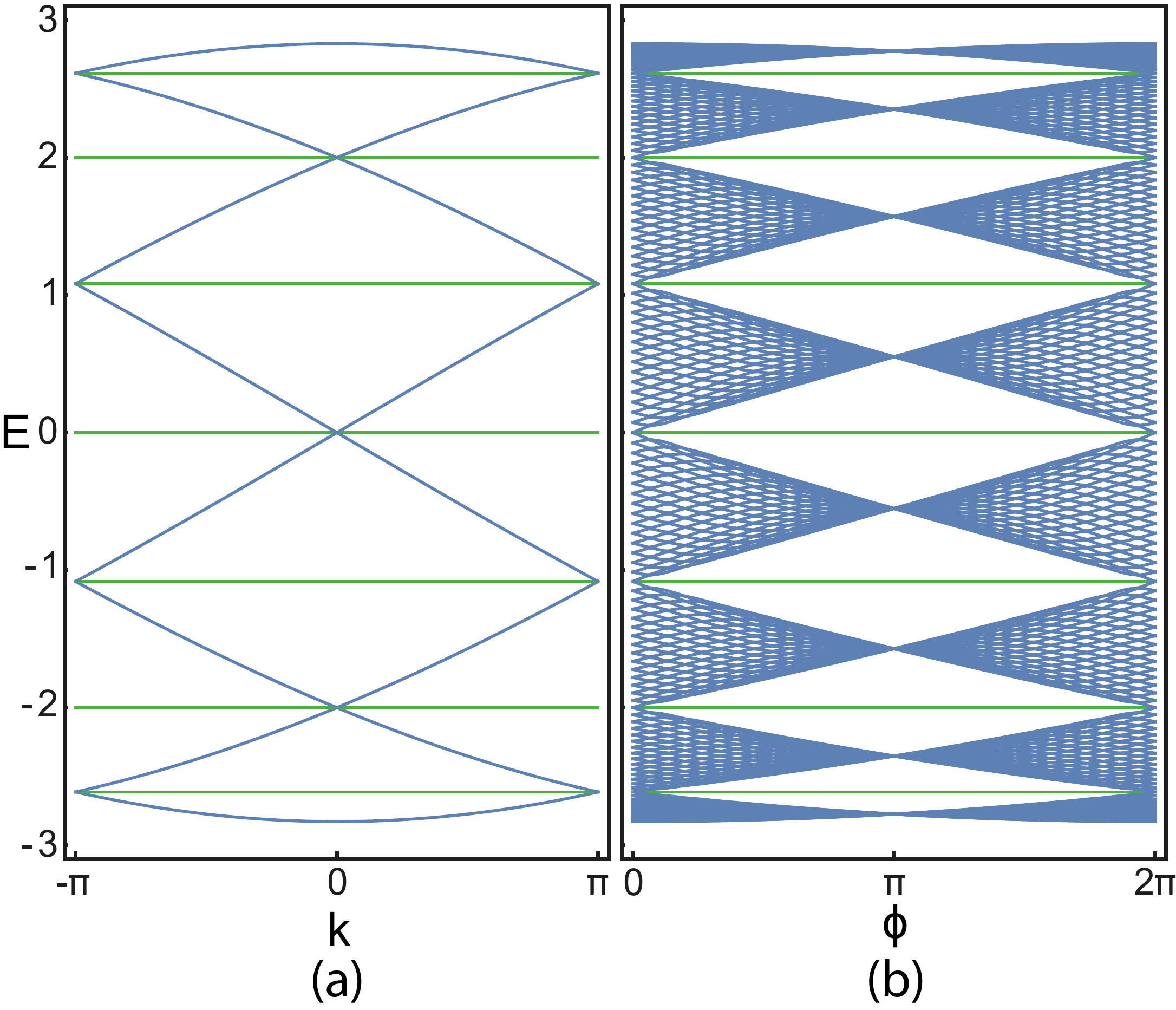}
		\par\end{centering}
	\caption{(a) Bulk energy spectrum of the $\sqrt[8]{\text{CL}}$.
		(b) Energy spectrum of a periodic $\sqrt[8]{\text{CL}}$ with $N_{uc}=30$ unit cells, as a function of the flux per plaquette.
		Dispersive (flat) bands are colored in blue (green).}
	\label{fig:3}
\end{figure}

\subsection{$2^n$-root topology}

The relation between the different root-degree versions of the $\sqrt[2^n]{\text{CL}}$ can be more easily understood by working with the $\mathcal{B}_B$ basis, where the two sublattices are separated.
The bulk Hamiltonian of the $\sqrt[2^n]{\text{CL}}$ has a block anti-diagonal form when written in $\mathcal{B}_B$,
\begin{equation}
H_{\sqrt[2^n]{\text{CL}},B}(k)=-\sqrt[2^n]{J}
\begin{pmatrix}
O_{2^{n}-1}&h_{\sqrt[2^n]{\text{CL}}}(k)
\\
h^\dagger_{\sqrt[2^n]{\text{CL}}}(k)&O_{2^{n}}
\end{pmatrix},
\label{eq:hamilcl2ton}
\end{equation}
where $h_{\sqrt[2^n]{\text{CL}}}(k)$, given in Appendix~\ref{app:hblocks}, is a $(2^{n}-1)\times 2^{n}$ matrix with the entire Peierls phase accumulated, as before, in the hopping between site $2^{n+1}-1$ and the spinal site 1.
The model has chiral-symmetry, defined as
\begin{eqnarray}
\mathcal{C}:\ \ 	&C&H_{\sqrt[2^n]{\text{CL}},B}(k)C^{-1}=-H_{\sqrt[2^n]{\text{CL}},B}(k),
	\\
	&C&=\text{diag}(I_{2^{n}-1},-I_{2^{n}}).
\end{eqnarray}
The square of this model reads as
\begin{eqnarray}
	H_{\sqrt[2^n]{\text{CL}},B}^2(k)&=&-\sqrt{2}
	\begin{pmatrix}
	H_{\sqrt[2^{n-1}]{\text{CL}},A}(k)&
	\\
	&H_{\sqrt[2^{n-1}]{\text{res},2^{n-1}}}(k)
	\end{pmatrix} \nonumber\\
 & & +c_{2^{n-1}}I_{2^{n+1}-1},\ \ n\geq 2,
	\label{eq:hsquaredcl}
	\end{eqnarray}
 \begin{eqnarray}
	H_{\sqrt[2^{n-1}]{\text{res},2^{n-1}}}(k)&=&\big(c_{2^{n-1}}I_{2^{n}}-h^\dagger_{\sqrt[2^n]{\text{CL}}}h_{\sqrt[2^n]{\text{CL}}}\big)/\sqrt{2},\nonumber\\
\end{eqnarray}
where the off-diagonal blocks in (\ref{eq:hsquaredcl}) are zero matrices, $c_{2^m}=4\sqrt[2^{m}]{J}$ is a constant energy shift and the minus sign was introduced in order to keep the sign convention for the hopping parameters as in (\ref{eq:hamilcl2ton}).
The squared Hamiltonian is block diagonal, with the top block $H_{\sqrt[2^{n}-1]{\text{CL}},A}(k)$ corresponding to the lower root-degree version of the model, occupying the blue sublattice and written in its $\mathcal{B}_A$ basis (since, in itself, the lower-root degree version is bipartite, with its own blue and green sublattices), and the lower block $H_{\sqrt[2^{n-1}]{\text{res},2^{n-1}}}(k)$ corresponding  to a residual chain with weight on the green sublattice only and a degenerate spectrum with the other block \cite{Ezawa2020,Marques2021,Marques2021b}, apart from an extra band with energy $E=c_{2^{n-1}}/\sqrt{2}$ coming from sublattice imbalance, in agreement with Lieb's theorem \cite{Lieb1989,Kikutake2013,Tindall2021,Marques2022}.

Upon changing the basis as $H_{\sqrt[2^{n}-1]{\text{CL}},A}(k)\to H_{\sqrt[2^{n}-1]{\text{CL}},B}(k)$, (\ref{eq:hsquaredcl}) can be iteratively applied until the diamond chain is reached, which has been shown to model a $\sqrt{\text{TI}}$ for finite flux that, in turn, derives its topological features from the topological block of its squared Hamiltonian, which describes a CL as the original TI  \cite{Kremer2020}.
Concretely, each of the $2^n$ edge states appearing for an open $\sqrt[2^n]{\text{CL}}$ has $1/2^n$ of its weight on the topological state of the CL \cite{Marques2021}.
In particular, when $\phi=\pi$, an all-bands-flat \cite{Danieli2020} bulk energy spectrum is found due to an AB caging effect \cite{Mukherjee2018,Pelegri2019,Pelegri2019b,Jorg2020}.
Recently, it has been shown that introducing modulated weak interactions in systems with an all-bands-flat spectrum, such as the diamond chain ($\sqrt{\text{CL}}$) \cite{Pelegri2020} and the CL \cite{Kuno2020b}, can induce the appearance of many-body topological states and interaction-driven dynamics \cite{Tovmasyan2013}.
In the opposite limit, that is, that of strong interactions, many-body subspaces exhibiting an all-bands-flat spectrum due to AB caging were shown to appear in a CL \cite{Nicolau2022}, even when the single-particle spectrum is dispersive.
\section{$2^n$-root square lattice}
\label{sec:2nrootsl}

\begin{figure}[tb]
	\begin{centering}
		\includegraphics[width=0.45 \textwidth]{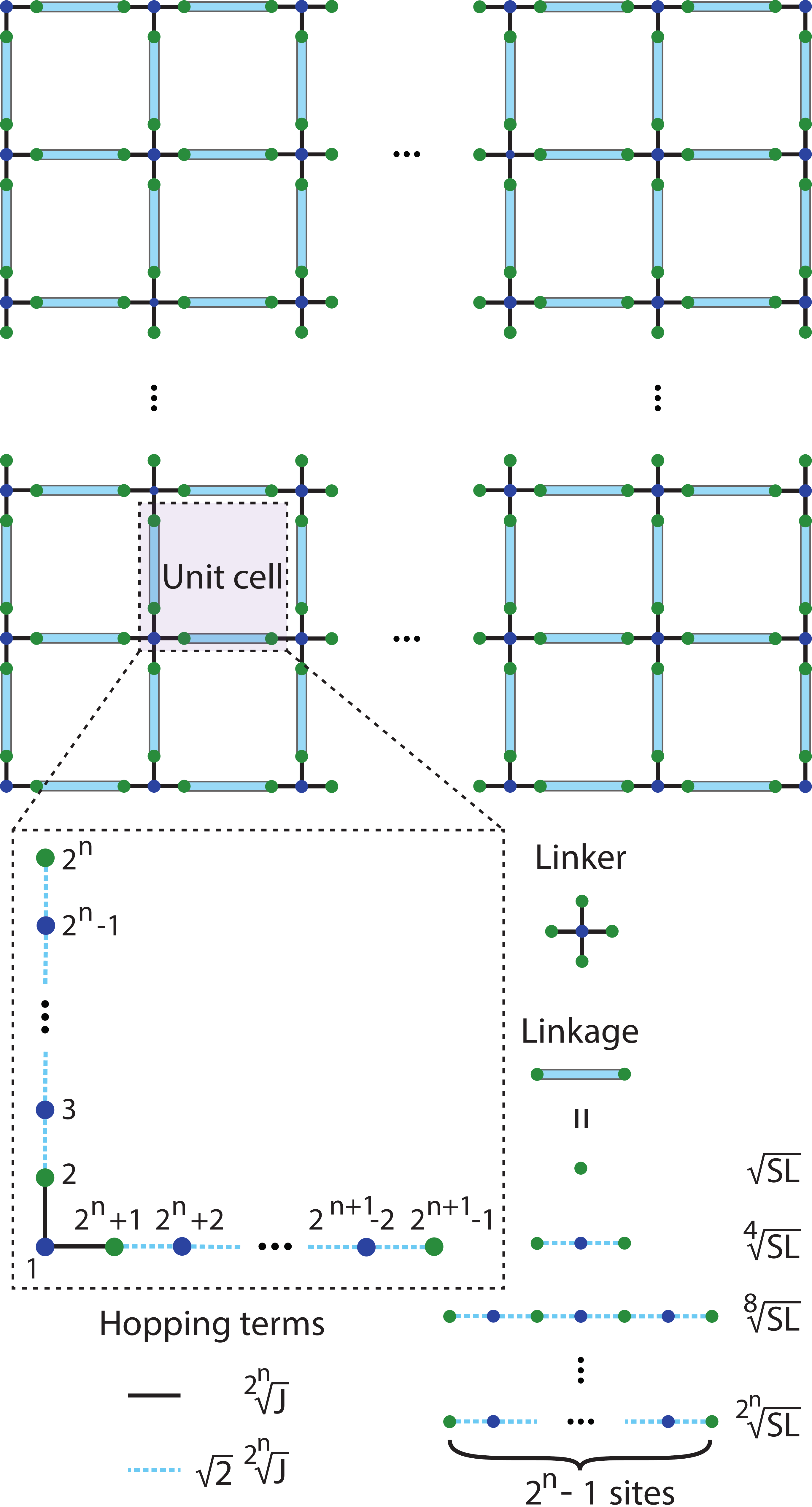}
		\par\end{centering}
	\caption{Illustration of the $\sqrt[2^n]{\text{SL}}$.
		The four green sites in the linker are shared with the adjacent linkages.
		The blue sublattice for a given $n$ constitutes both sublattices of the $n-1$ chain.}
	\label{fig:4}
\end{figure}

We turn now our attention to a family of lattices that can be understood as the 2D generalization of the $\sqrt[2^n]{\text{CLs}}$ analyzed above, namely the family of $\sqrt[2^n]{\text{SL}}$ models depicted in Fig.~\ref{fig:4}.
It is clear that the $\sqrt{\text{SL}}$ is nothing more than the well-known Lieb lattice, a correspondence that will be further explored below.
In fact, following recent interest in decorated Lieb lattices \cite{Morales2016,Rontgen2019,Bhattacharya2019,Ni2020}, the $\sqrt[2^n]{\text{SL}}$ can be viewed as a specific type of Lieb-$(2^{n+1}-1)$ lattice \cite{Zhang2017,Jiang2019}, where $2^{n+1}-1$ is the number of sites per unit cell, also labeled $\mathcal{L}(2^n-1)$ \cite{Mao2020}, where the argument is the number of sites in the linkage.
As before, we will distinguish between the two bases $\mathcal{B}_A$ and $\mathcal{B}_B$ for the Hilbert space defined in (\ref{eq:basisa})-(\ref{eq:basisb}),
with BS (GS) labeling now the ordered sites within the blue (green) sublattice of the unit cell of the $\sqrt[2^n]{\text{SL}}$ in Fig.~\ref{fig:4}.

\begin{figure}[ht]
	\begin{centering}
		\includegraphics[width=0.45 \textwidth]{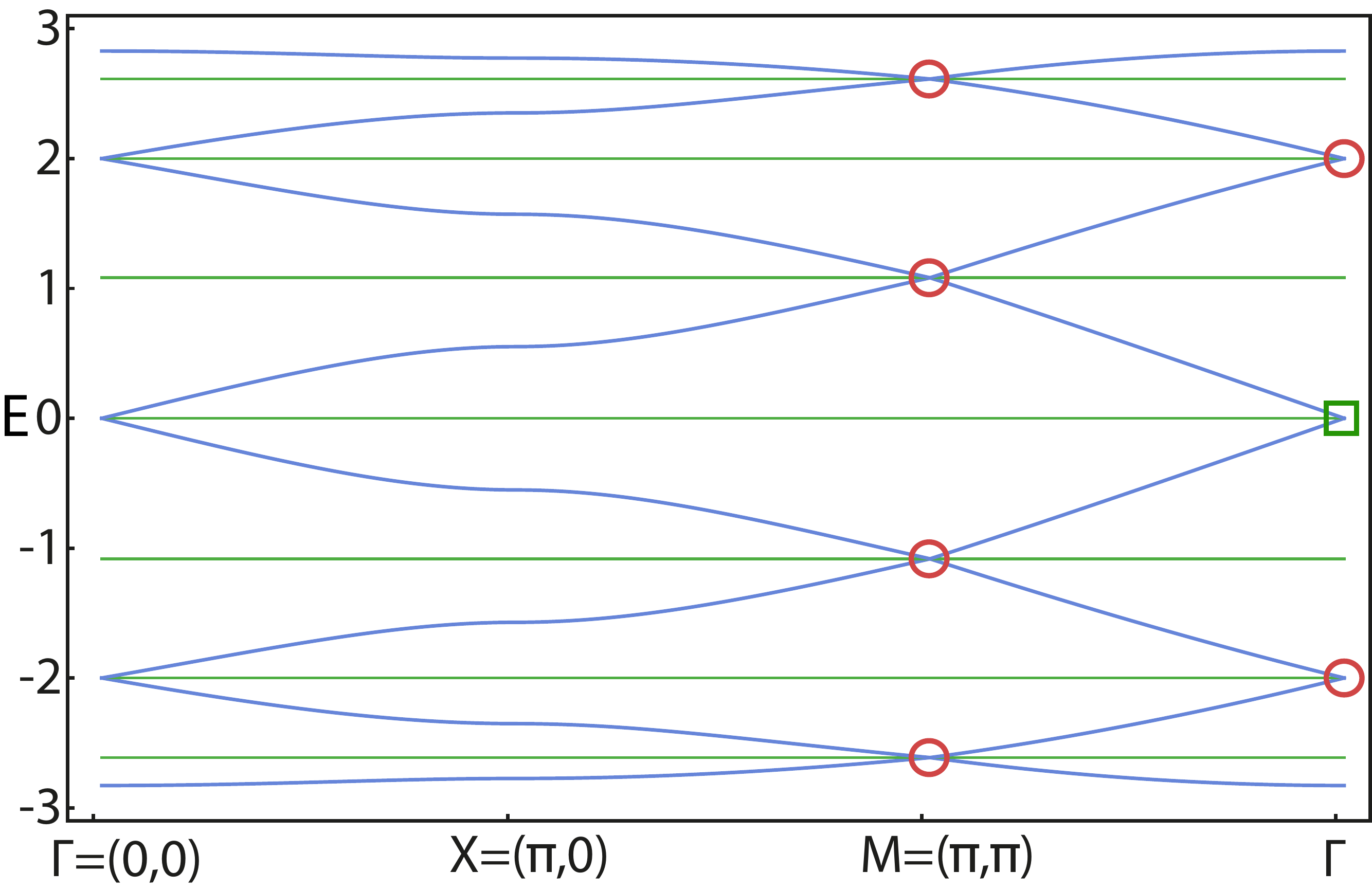}
		\par\end{centering}
	\caption{Bulk energy spectrum of the $\sqrt[8]{\text{SL}}$ along the high-symmetry lines of the Brillouin zone.
		An open green square (red circle) indicates a particle-hole symmetric (asymmetric) spin-1 Dirac cone centered at a high-symmetry point.}
	\label{fig:5}
\end{figure}

In the $\mathcal{B}_A$ basis, the bulk Hamiltonian of the $\sqrt[2^n]{\text{SL}}$ without flux and lattice constants set to $a_x=a_y=1$ is written as
	\begin{equation}
	H_{\sqrt[2^n]{\text{SL}},A}(\mathbf{k})=
	\begin{pmatrix}
		0&\mathbf{u}_{2^n}(k_y)&\mathbf{u}_{2^n}(k_x)
		\\
		\mathbf{u}^{\dagger}_{2^n}(k_y)&H^{\text{SL}}_{\text{Linkage},2^n}&O_{2^{n}-1}
		\\
		\mathbf{u}^{\dagger}_{2^n}(k_x)&O_{2^{n}-1}&H^{\text{SL}}_{\text{Linkage},2^n}&
	\end{pmatrix},
	\label{eq:hamilt2nsl}
\end{equation}
where $\mathbf{u}_{2^n}(k_\alpha)=-\sqrt[2^n]{J}(1,0,\dots,0,e^{-ik_\alpha})$ is a vector of size $2^{n}-1$ and $H^{\text{SL}}_{\text{Linkage},2^n}$ is the tridiagonal Hamiltonian of an open linear chain of $2^{n}-1$ sites and uniform nearest-neighbor couplings $\sqrt{2}\sqrt[2^n]{J}$.
As an example, the bulk energy spectrum of the $\sqrt[8]{\text{SL}}$ along the high-symmetry lines of the Brillouin zone is shown in Fig.~\ref{fig:5}, which has similar features to that of the $\sqrt[4]{\text{CL}}$ in Fig.~\ref{fig:3}(a), particularly with regard to the formation of seven flat bands, placed now at the eigenvalues of $H^{\text{SL}}_{\text{Linkage},8}$ in (\ref{eq:hamilt2nsl}).
Another remarkable feature of these models, as highlighted in Fig.~\ref{fig:5}, is the appearance of $2^n-1$ spin-1 Dirac cones in the spectrum, with a particle-hole symmetric one centered at zero-energy and $2^n-2$ particle-hole asymmetric others centered at either the $\Gamma$ or $M$ point of the finite energy flat bands.

In order to prove that the energies of the flat bands of the $\sqrt[2^n]{\text{SL}}$ correspond to the eigenergies of $H^{\text{SL}}_{\text{Linkage},2^n}$ one needs, similarly to (\ref{eq:intertwiner}), to find an intertwiner that satisfies
\begin{equation}
	H_{\sqrt[2^n]{\text{SL}},A}(\mathbf{k})G_{2^n}^\prime(\mathbf{k})=G_{2^n}^\prime(\mathbf{k})H^{\text{SL}}_{\text{Linkage},2^n}.
	\label{eq:intertwinerprime}
\end{equation}
It can be readily checked that the intertwiner can be written in this case as
\begin{equation}
	G_{2^n}^\prime(\mathbf{k})=\begin{pmatrix}
		\vec{O}_{2^{n}-1}
		\\
		\alpha^\prime I_{2^{n}-1}+\gamma^\prime J_{2^{n}-1}
		\\
		\beta^\prime I_{2^{n}-1}+\delta^\prime J_{2^{n}-1}
		\label{eq:intertwinersl}
	\end{pmatrix},
\end{equation}
with the coefficients given by
\begin{eqnarray}
	\gamma^\prime&=&\delta^\prime:=e^{-ik_x}-e^{-ik_y},
	\\
	\alpha^\prime&=&2-e^{-2ik_x}-e^{-i(k_x+k_y)},
	\\
	\beta^\prime&=&e^{-i(k_x+k_y)}+e^{-2ik_y}-2.
\end{eqnarray}
Then, the eigenvalues of the linkage, whose correspondent eigenstates satisfy
\begin{eqnarray}
	H^{\text{SL}}_{\text{Linkage},2^n}\varphi^\prime_{2^n,j}&=&\epsilon_j\varphi^\prime_{2^n,j},
	\label{eq:schroedlinkagesl}
	\\
	\epsilon_j&=&-2\sqrt{2}\cos(\frac{\pi}{2^{n}}j),
	\label{eq:enerslflat}
\end{eqnarray}
with $ j=1,2,\dots,2^{n}-1$ and $\varphi^\prime_{2^n,j}\not\in\ker G_{2^n}^\prime(\mathbf{k})$ assumed, are also eigenvalues of $H_{\sqrt[2^n]{\text{SL}},A}(\mathbf{k})$, following the same reasoning as in (\ref{eq:commoneigs}).
Finally, the $j^\text{th}$ eigenstate within the flat band subspace of the bulk Hamiltonian $H_{\sqrt[2^n]{\text{SL}},A}(\mathbf{k})$ can be written as
\begin{equation}
	\psi^\prime_{2^n,j}(\mathbf{k})=\frac{1}{\mathcal{N}_{2^n}^{j\prime}(\mathbf{k})}G_{2^n}^\prime(\mathbf{k})\varphi^\prime_{2^n,j},
	\label{eq:bulkflateigssl}
\end{equation}
with $\mathcal{N}_{2^n}^{j\prime}(\mathbf{k})$ the normalization constant.

\begin{figure}[ht]
	\begin{centering}
		\includegraphics[width=0.49 \textwidth]{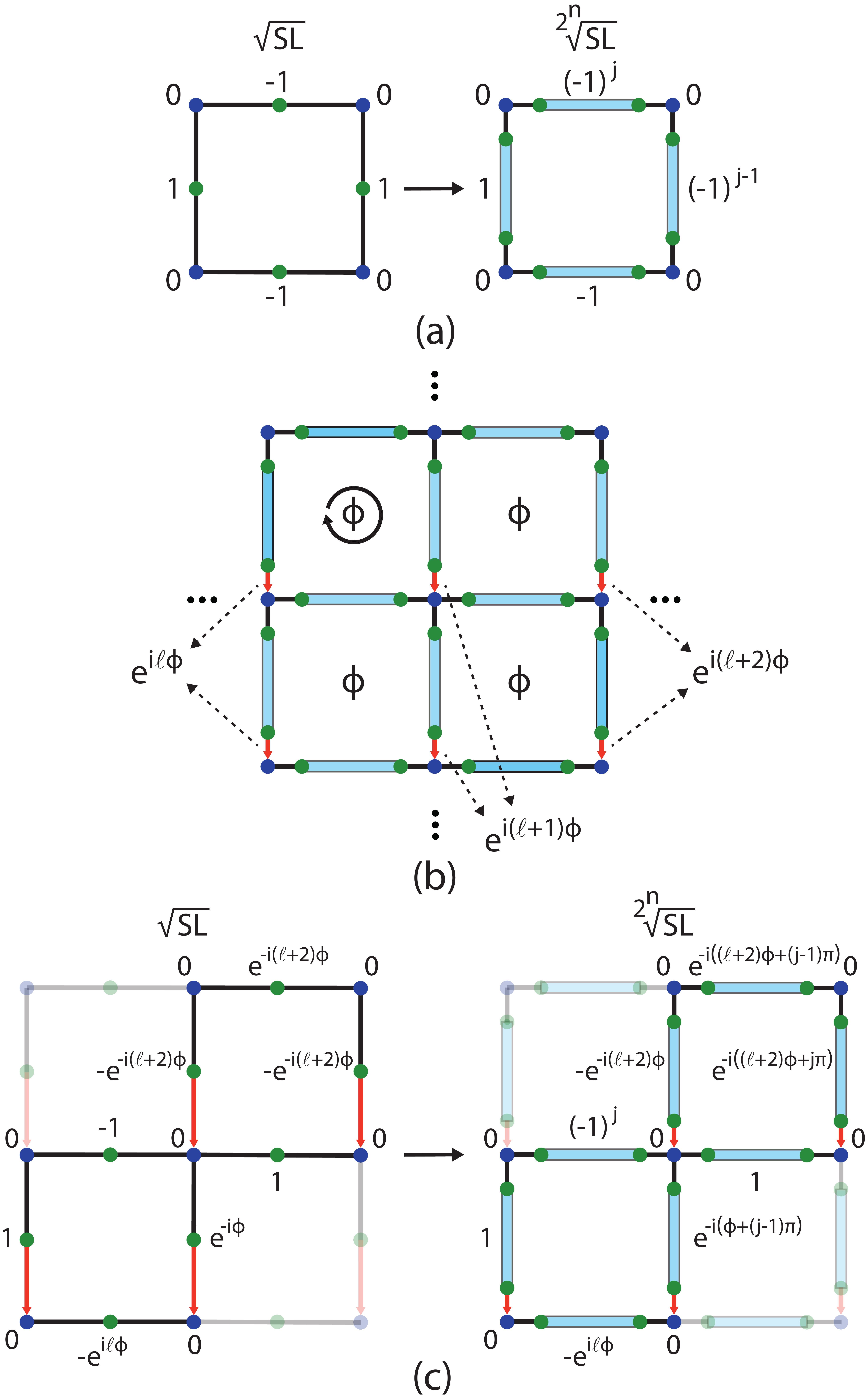}
		\par\end{centering}
	\caption{(a) Map of the zero-energy CLS in the $\sqrt{\text{SL}}$ (Lieb lattice), for zero flux, onto the CLS of the $j^{th}$ flat band of the $\sqrt[2^n]{\text{SL}}$, with $n\geq 2$ and $j=1,2,\dots,2^n-1$ the index of the linkage eigenstate in (\ref{eq:schroedlinkagesl}).
		The components at the green sites at the left translate as global phases for the linkages at the right, dependent on the oddness of the $j^{th}$ harmonic.
		(b) Gauge choice, away from the boundaries, for the introduction of a $\phi$ reduced flux per plaquette in the $\sqrt[2^n]{\text{SL}}$, where only the red couplings have finite Peierls phases, positively accumulated along the direction of their arrows, and $l\in \mathbb{N}$.
		(c) Same as in (a), but for a finite $\phi$ flux per plaquette.}
	\label{fig:6}
\end{figure}

The form of the CLS of each flat band of the $\sqrt[2^n]{\text{SL}}$ cannot be found directly with the intertwiner $G_{2^n}^\prime(\mathbf{k})$ in (\ref{eq:intertwinersl}), since its irremovable $\mathbf{k}$-dependence indicates that the CLSs span more than one unit cell.
However, starting from the well-know form of a CLS in the Lieb lattice (the $\sqrt{\text{SL}}$), which constitutes a chiral CLS of a linear flat band Hamiltonian \cite{Graf2021}, it is straightforward to generalize it in order to find all CLSs of the $\sqrt[2^n]{\text{SL}}$.
This generalization is schematically depicted in Fig.~\ref{fig:6}(a) where, from the CLS of the Lieb lattice at the left, occupying a single plaquette by virtue of having nodes at the blue sites connecting it to the adjacent ones, the CLS of flat band $j=1,2,\dots,2^n-1$ of the $\sqrt[2^n]{\text{SL}}$, with an energy given in (\ref{eq:enerslflat}), is directly inferred at the right.
The finite components at the right correspond to a global phase factor applied to the $j^{th}$-harmonic of the respective linkage [see $\varphi^\prime_{2^n,j}$ in (\ref{eq:schroedlinkagesl})], with site index increasing from bottom to top for the vertical linkages and from left to right on the horizontal linkages.
While the phase factors for odd $j$ are the same as for the Lieb lattice, the top and right ones are inverted for even $j$, due to the fact that even harmonics have opposite signs at their end sites.
Note, however, that the CLSs cannot be used to define an orthonormal basis, since each CLS has a finite overlap with the CLSs lying in the adjacent plaquettes.

\subsection{Hierarchy of different root-degree versions}

Turning now to the manifestly chiral-symmetric $\mathcal{B}_B$ basis, the bulk Hamiltonian of the $\sqrt[2^n]{\text{SL}}$ can be rewritten as
	\begin{equation}
		H_{\sqrt[2^n]{\text{SL}},B}(\mathbf{k})=-\sqrt[2^n]{J}
		\begin{pmatrix}
			O_{2^{n}-1}&h_{\sqrt[2^n]{\text{SL}}}(\mathbf{k})
			\\
			h^\dagger_{\sqrt[2^n]{\text{SL}}}(\mathbf{k})&O_{2^{n}}
		\end{pmatrix},
		\label{eq:hamilsl2ton}
	\end{equation}
where $h_{\sqrt[2^n]{\text{SL}}}(\mathbf{k})$ is a $(2^{n}-1)\times 2^{n}$ matrix explicitly given in Appendix~\ref{app:hblocks}.
The chiral-symmetry is defined as
\begin{eqnarray}
	\mathcal{C}:\ \ 	&C&H_{\sqrt[2^n]{\text{SL}},B}(\mathbf{k})C^{-1}=-H_{\sqrt[2^n]{\text{SL}},B}(\mathbf{k}),
	\\
	&C&=\text{diag}(I_{2^{n}-1},-I_{2^{n}}).
\end{eqnarray}
Upon squaring the Hamiltonian in (\ref{eq:hamilsl2ton}) one arrives at
\begin{widetext}
	\begin{eqnarray}
		H_{\sqrt[2^n]{\text{SL}},B}^2(\mathbf{k})&=&c_{2^{n-1}}^\prime I_{2^{n+1}-1}-\sqrt{2}
		\begin{pmatrix}
			H_{\sqrt[2^{n}-1]{\text{SL}},A}(\mathbf{k})&
			\\
			&H^\prime_{\sqrt[2^{n-1}]{\text{res},2^{n-1}}}(\mathbf{k})
		\end{pmatrix},\ \ n>1,
		\label{eq:hsquaredsl}
		\\
		H^\prime_{\sqrt[2^{n-1}]{\text{res},2^{n-1}}}(\mathbf{k})&=&\big(c_{2^{n-1}}^\prime I_{2^{n+1}}-h^\dagger_{\sqrt[2^n]{\text{SL}}}h_{\sqrt[2^n]{\text{SL}}}\big)/\sqrt{2},
		\label{eq:hsquaredslres}
	\end{eqnarray}
\end{widetext}
where the off-diagonal blocks in (\ref{eq:hsquaredsl}) are zero matrices, $c_{2^m}^\prime=4\sqrt[2^{m}]{J}$ is a constant energy shift and the minus sign was introduced in order to keep the sign convention for the hopping parameters as in (\ref{eq:hamilsl2ton}).
The squared Hamiltonian is block diagonal, with the top block $H_{\sqrt[2^{n}-1]{\text{SL}},A}(\mathbf{k})$ corresponding to the lower root-degree version of the SL, occupying the blue sublattice and written in its $\mathcal{B}^\prime_A$ basis, and the lower block $H^\prime_{\sqrt[2^{n-1}]{\text{res},2^{n-1}}}(\mathbf{k})$ corresponding  to a residual lattice with weight on the green sublattice only and a degenerate spectrum with the other block \cite{Ezawa2020,Marques2021,Marques2021b}, apart from an extra band with energy $E=c_{2^{n-1}}^\prime/\sqrt{2}$ coming from sublattice imbalance, corresponding to the $E=0$ band of the starting $H_{\sqrt[2^n]{\text{SL}},B}(\mathbf{k})$.
Note that the same relations hold for $n=1$ (squaring the Lieb lattice to arrive at the SL in one of the blocks) upon dropping the $\sqrt{2}$ factor.
After changing the basis as $H_{\sqrt[2^{n}-1]{\text{SL}},A}(\mathbf{k})\to H_{\sqrt[2^{n}-1]{\text{SL}},B}(\mathbf{k})$, (\ref{eq:hsquaredsl}) can be iteratively applied until the SL Hamiltonian is reached and, in the process, $n$ additional residual lattices with degenerate spectra (apart from bands originating from sublattice imbalance) will appear, assuming their Hamiltonians are squared, shifted in energy by $c_{2^m}^\prime$, with successive $m=n-1,n-2,\dots,0$ \cite{Marques2021,Marques2021b}, and renormalized by $-1/\sqrt{2}$ in parallel at each step, except for the last one where no renormalization is needed.

Due to the hierarchy of root-degree versions of our model, connecting the $\sqrt[2^n]{\text{SL}}$ to the SL by $n$ successive applications of the squaring operation in (\ref{eq:hsquaredsl}), whose process was seen elsewhere to generate a rooted tree graph directed from the former to the latter \cite{Marques2021}, the energies of the dispersive bands of the $\sqrt[2^n]{\text{SL}}$ can be compactly written as
\begin{multline}
	\epsilon^{\sqrt[2^n]{\text{SL}}}_{\underbrace{\pm\pm \pm \cdots \pm\pm}_{n}}(\mathbf{k})= \\
	\pm \sqrt{c_{2^n}^\prime\pm\sqrt{2}\sqrt{c_{2^{n-1}}^\prime\pm\sqrt{2}\sqrt{\ddots \sqrt{2}\sqrt{c_1^\prime - \epsilon^{\text{SL}}(\mathbf{k})}}}},
	\label{eq:enersldispersive}
\end{multline}
with $\epsilon^{\text{SL}}(\mathbf{k})=-2J(\cos k_x + \cos k_y)$ the energy band of the SL.
Setting $J\equiv1$ simplifies all constant energy shifts to $c_{2^m}^\prime\to 4$.
This type of compact notation was used by some authors in the context of certain linear 1D models, labeled sine-cosine models, exhibiting self-similar properties upon successive squaring operations \cite{Dias2021}.
The full ($2^{n+1}-1$)-bands energy spectrum of the $\sqrt[2^n]{\text{SL}}$ results from the combination of the $2^n$ dispersive bands coming from (\ref{eq:enersldispersive}) with the $2^n -1$ energy bands within the flat band subspace found in (\ref{eq:enerslflat}).

\section{Hofstadter spectrum of the $\sqrt[2^n]{\text{SL}}$}
\label{sec:hofst}

\begin{figure*}[ht]
	\begin{centering}
		\includegraphics[width=0.95 \textwidth,height=13cm]{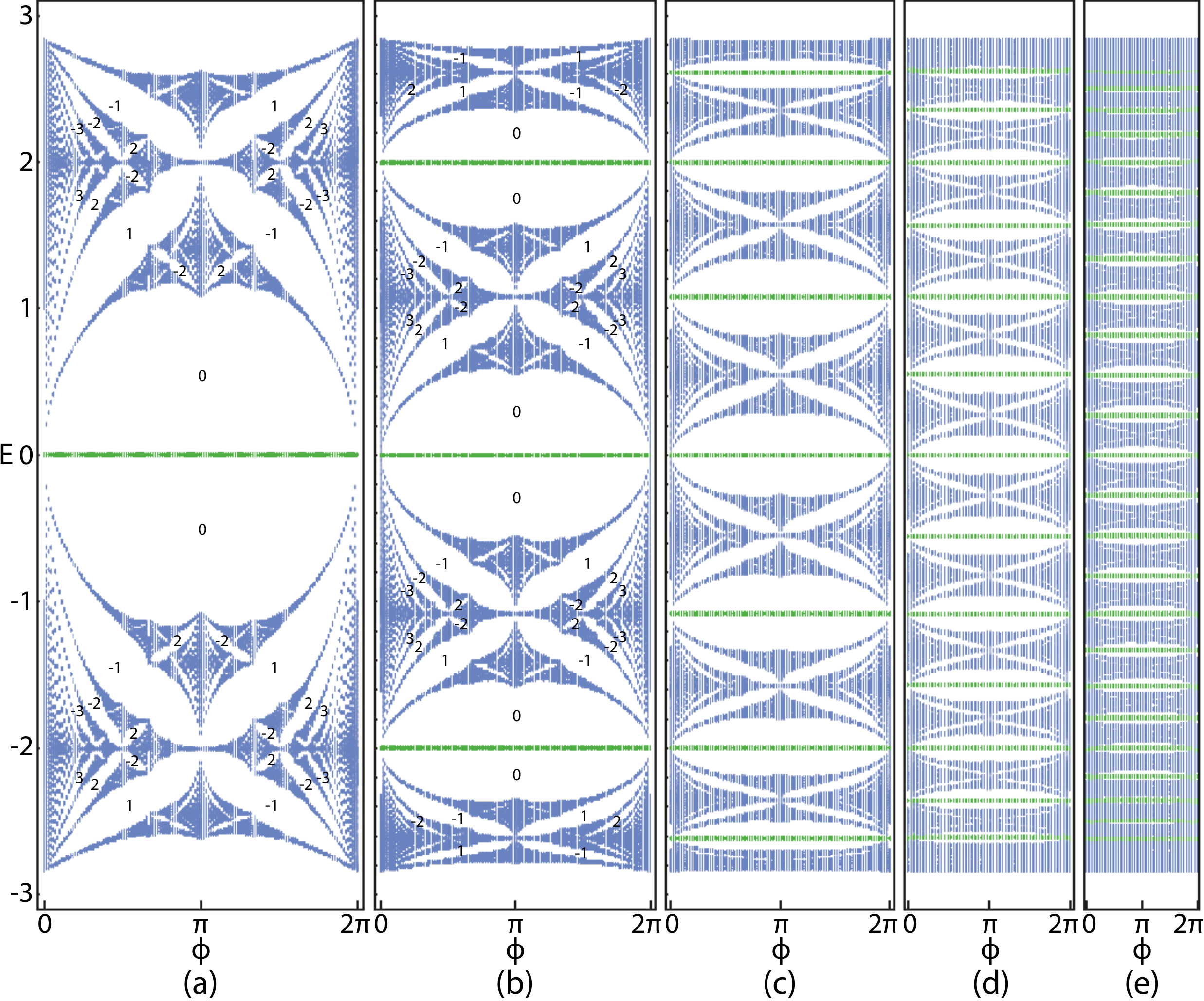}
		\par\end{centering}
	\caption{ Hofstadter spectrum of the periodic $\sqrt[2^n]{\text{SL}}$ scanned over the rational fluxes $\phi=2\pi \frac{l}{q}$, with $l=0,1,\dots,q$, for (a) $n=1$ and $q=141$, (b) $n=2$ and $q=131$, (c) $n=3$ and $q=83$, (d) $n=4$ and $q=53$, and (e) $n=5$ and $q=41$.
		Each spectrum displays $2^n$ blue butterflies intercalated with $2^{n}-1$ flux insensitive green flat bands.
		The Chern numbers at some representative gaps are indicated in (a) and (b).}
	\label{fig:7}
\end{figure*}
In this section, we will study the behavior of the $\sqrt[2^n]{\text{SL}}$ when a $\phi$ magnetic flux per plaquette is applied.
The gauge choice for the introduction of the flux is shown in Fig.~\ref{fig:6}(b), where only the red vertical hoppings have a finite Peierls phase.
If periodic boundary conditions are applied in one or both directions, modifications to the gauge configuration along the boundary plaquettes are in order which, however, do not change the physical picture.
As such, we will focus on the bulk plaquettes when discussing the CLSs below, which applies both to OBC and PBC.

\begin{figure}[tbh]
	\begin{centering}
		\includegraphics[width=0.45 \textwidth]{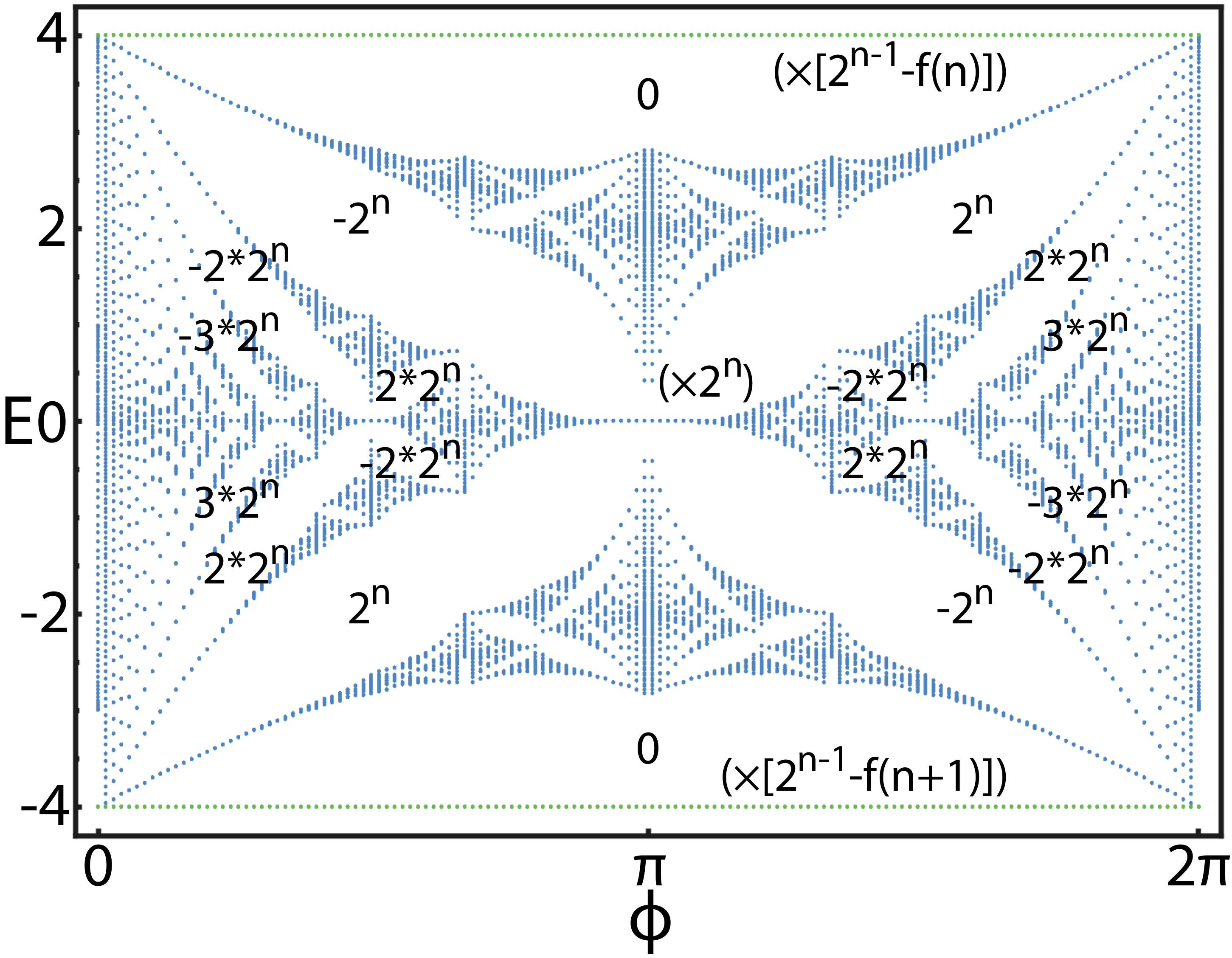}
		\par\end{centering}
	\caption{Hofstadter spectrum of the $n$-times squared $\sqrt[2^n]{\text{SL}}$, with a global energy downshifting and renormalization after each squaring operation, as detailed in Sec.~\ref{sec:2nrootsl}A.
		The expressions in parentheses indicate the global degeneracy of the top and bottom green flat bands placed at $E=\pm c_1$ and of the middle blue butterfly, with $f(n)=0(1)$ for $n$ even (odd).
		The Chern numbers at some representative gaps are indicated.
		}
	\label{fig:8}
\end{figure}

In Fig.~\ref{fig:7}, we show the Hofstadter spectrum of the $\sqrt[2^n]{\text{SL}}$, for $n=1,\dots,5$, scanned over the rational reduced fluxes $\phi=2\pi\frac{l}{q}=2\pi \frac{\Phi}{\phi_0}$, with $\phi_0$ the flux quantum, $q$ a prime number, $l=0,1,\dots,q$, $\Phi=\frac{l}{q}\phi_0$ the magnetic flux per plaquette and assuming periodic boundary conditions along both directions, with $N_x=3$ unit cells in the $x$ direction and $N_y=q$ unit cells in the $y$ direction.
For $n=1$, in Fig.~\ref{fig:7}(a), we recover the Hofstadter spectrum of the Lieb lattice, displaying two symmetric butterflies separated by a zero-energy flat band \cite{Aoki1996,Goldman2011,Nita2013,Xu2018}.
Its relation to the spectrum of the SL is revealed in Fig.~\ref{fig:8}, which corresponds to squaring the spectrum of Fig.~\ref{fig:7}(a) with a global downshift of $c_1^\prime=4J$, that is plotting $E_{\sqrt{\text{SL}}}^2(\phi)-c_1^\prime$.
As also shown in the bottom panel of Fig.~\ref{fig:1}, this spectrum can be decomposed in a butterfly coming from the SL diagonal block on the blue sublattice one arrives at by squaring the $\sqrt{\text{SL}}$ Hamiltonian, while both the other degenerate butterflies and the flat band at $E=-c_1^\prime$ are originated by diagonalization of the residual block built on the green sublattice.
The one-time squared Hofstadter spectrum of Fig.~\ref{fig:8} confirms that the Lieb lattice can indeed be understood as the square-root version of the SL.
In a similar fashion, the Hofstadter spectrum of the honeycomb-kagome or superhoneycomb lattice (see Fig.~4(a) of [\onlinecite{Aoki1996}]) displays the same qualitative features as the Lieb lattice in Fig.~\ref{fig:7}(a), namely a pair of symmetric butterflies separated by a zero-energy flat band which, when squared, retrieves the doubly degenerate spectrum of the honeycomb lattice, limited below by the flat band.
Thus, the superhoneycomb lattice is the square-root version of the honeycomb lattice, as was also highlighted in recent studies \cite{Mizoguchi2020,Mizoguchi2021c,Marques2021b}.

Concerning the spectra in Figs.~\ref{fig:7}(b)-(e), as $n$ increases a kaleidoscope of $2^n$ butterflies is seen to appear, each separated by the adjacent ones by a flux insensitive flat band, which form a subspace of $2^{n}-1$ flat bands adiabatically connected to the flat band subspace at zero flux [compare the $n=3$ case of Fig.~\ref{fig:7}(c), for $\phi =0$, with the corresponding bulk spectrum in Fig.~\ref{fig:5}].
It should be stressed that, to the best of our knowledge, the $\sqrt[2^n]{\text{SL}}$ with $n\geq 2$ is the first example of a lattice (or family of different root-degree versions of the same model, the SL) exhibiting multiple flux insensitive flat bands in its Hofstadter diagram, generalizing for two-dimensional (2D) systems the behavior found in the diagram of the quasi-1D $\sqrt[2^n]{\text{CL}}$ for $n\geq 1$ [see Fig.~\ref{fig:3}(b) for the $n=3$ case].
In our view, the main reason for this gap in the literature is that most systems with multiple flat bands in their bulk spectrum, that is, for zero flux, see the flat band states go from localized to extended when a finite magnetic flux is applied, rendering the flat bands dispersive.
In contrast, we will show in the next section that, for a finite flux per plaquette in the $\sqrt[2^n]{\text{SL}}$, two plaquettes sharing a corner site can serve as a common AB cage that is able to accommodate the CLSs of all the flat bands in a locally orthogonal basis, such that these survive the introduction of flux.

\subsection{Topological characterization of the $\sqrt[2^n]{\text{SL}}$}

 The relevant strong topological index at each gap of the Hofstadter spectrum is the Chern number \cite{Hasan2010}, which can be computed through the Streda formula \cite{Streda1982},
\begin{equation}
	C=\phi_0\frac{d\rho}{dB}\bigg|_{E=E_F},
	\label{eq:chernsmooth}
\end{equation}
where $E_F$ is the Fermi level placed inside the considered energy gap, $\rho=N_{\text{gap}}/A$ is the density of occupied states, with $N_{\text{gap}}$ the total number of occupied states and $A=N_x a_x q a_y$ the area of the lattice (since we set $N_y=q$), and $B$ the magnetic flux.
Given that we are restricted to a discrete grid of rational magnetic fluxes per unit cell, $\Phi=B a_x a_y=\frac{l}{q}\phi_0$ with $l=0,1,\dots,q-1$, it is convenient to work with a discretized version \cite{Fukui2016,Mizoguchi2021} of (\ref{eq:chernsmooth}),
\begin{equation}
	C=\phi_0\frac{\Delta N_{\text{gap}}}{A\Delta B}=2\pi\frac{\Delta N_{\text{gap}}}{N_x q\Delta\phi},
\end{equation}
where the $2\pi$ factor comes from the definition of the reduced flux, $\phi=2\pi\frac{\Phi}{\phi_0}$, and $\Delta N_{\text{gap}}$ is the variation of the number of occupied states as the flux is varied by $\Delta\phi=\frac{2\pi}{q}$, which leads finally to the simple formula
\begin{equation}
	C=\frac{\Delta N_{\text{gap}}}{N_x}.
	\label{eq:cherndiscrete}
\end{equation}

Using (\ref{eq:cherndiscrete}), the values for the Chern number at some representative gaps is shown for the Lieb lattice \cite{Goldman2011} in Fig.~\ref{fig:7}(a), where it can be seen that each butterfly reproduces the well-know behavior of the butterfly of the SL \cite{Satija2016} at equivalent energy gaps.
The same behavior is observed for each butterfly on the Hofstadter spectrum of the $\sqrt[2^n]{\text{SL}}$ for all $n$.
In Fig.~\ref{fig:7}, the Chern number at the gaps between the flat bands and the butterflies connected from above and below at $\phi=0$ is zero for all cases.
Focusing on the Lieb lattice, both these results can be straightforwardly derived as follows:
\\\\
(i) Apart from the flat band, the two diagonal blocks in $H^2_{\sqrt{\text{SL}}}$ share the same spectrum \cite{Ezawa2020,Marques2021,Marques2021b}.
As such, both the top and bottom butterflies in Fig.~\ref{fig:7}(a) square to the Hofstadter butterfly and, thus, share with it the same topological characterization.
The same reasoning can be extended to all $\sqrt[2^n]{\text{SL}}$ where, after $n$ squaring, energy downshifting and renormalization operations, as detailed in Sec.~\ref{sec:2nrootsl}A, one arrives at a $2^n$-fold degenerate Hofstadter butterfly, as illustrated in Fig.~\ref{fig:8}.
One can also conclude from this degeneracy that all successive residual lattice models are Chern insulators.
\\\\
(ii) Since, on the one hand, the Chern numbers at the gaps of any given butterfly sum to zero at every $\phi$ and, on the other hand, it has been proven \cite{Chen2014} that flat bands in systems with local hoppings, as is our case, have trivial Chern numbers (or total Chern numbers for degenerate bundles), it follows that all the gaps between flat bands and butterflies in the $\sqrt[2^n]{\text{SL}}$ have trivial Chern numbers.
\\\\
\begin{figure}[tb]
	\begin{centering}
		\includegraphics[width=0.49 \textwidth]{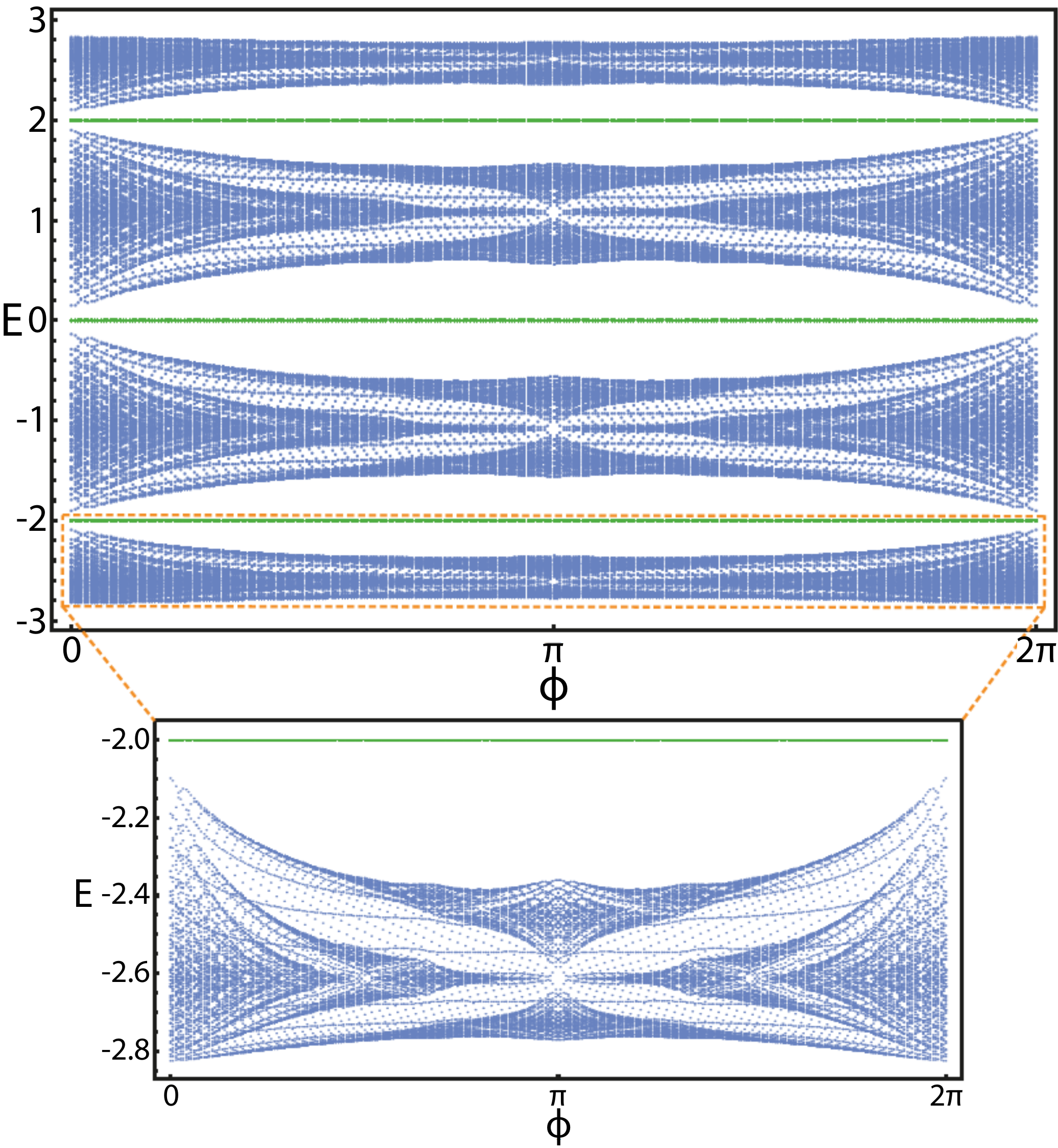}
		\par\end{centering}
	\caption{Hofstadter spectrum of the $\sqrt[4]{\text{SL}}$ depicted in Fig.~\ref{fig:4} with $10\times 10$ plaquettes, periodic in $x$ and open in $y$.
		The inset shows a zoomed version of the bottom flat band and butterfly.
		The energy gaps between the flat bands and the butterflies at $\phi = 0, 2 \pi$ are an artifact of the finite size of the lattice.}
	\label{fig:9}
\end{figure}
As an example, the Hofstadter spectrum of the $\sqrt[4]{\text{SL}}$ with $10\times 10$ plaquettes and periodic (open) boundaries in $x$ ($y$) is plotted in Fig.~\ref{fig:9}.
The edge states are seen to appear only at the gaps within each butterfly, in agreement with the Chern numbers of Fig.~\ref{fig:7}(b), and the flat bands are confirmed to be inert with regard to Chern topology.
A comparison between the lower butterfly zoomed in on the inset and the two middle ones highlights that, for each $\phi$, the number of edge states is the same at equivalent gaps of the butterflies, since they all share the same topological features.

\subsection{Multiple Aharonov-Bohm caging}

\begin{figure}[tb]
	\begin{centering}
		\includegraphics[width=0.48 \textwidth]{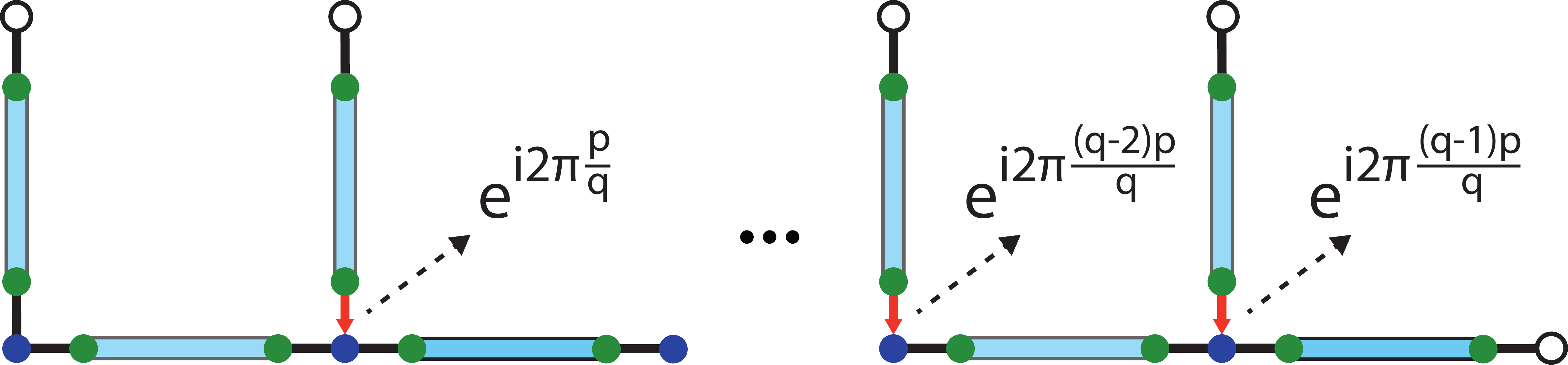}
		\par\end{centering}
	\caption{Magnetic unit cell of the $\sqrt[2^n]{\text{SL}}$ with a flux $\phi=2\pi\frac{p}{q}$ per plaquette, with $p$ and $q$ coprime integers, consisting of $q$ structural unit cells.
		The form of the light blue linkages for different $n$ and the values of the hopping parameters are the same as in Fig.~\ref{fig:4}, while the gauge configuration follows the scheme indicated in Fig.~\ref{fig:6}(b).
		Open sites at the edges belong to adjacent unit cells.}
	\label{fig:10}
\end{figure}

Contrary to the $\sqrt[2^n]{\text{CL}}$ studied in Sec.~\ref{sec:2ncl}, where the magnetic and structural unit cells were of the same size, the magnetic unit cell of the $\sqrt[2^n]{\text{SL}}$ can only be defined for rational values of the flux and is larger than the structural unit cell depicted in Fig.~\ref{fig:4}.
Furthermore, its size varies with the flux value considered, as shown in Fig.~\ref{fig:10}.
However, the analytical expression for the flat band eigenstates, which relies on finding a generalized version of the intertwiner in (\ref{eq:intertwinersl}), can still be derived from the bulk Hamiltonian, as we demonstrate in Appendix~\ref{app:intertwinerslflux}.

Here we are interested, on the other hand, in the question of whether the flat band eigenstates of the $\sqrt[2^n]{\text{SL}}$ can be analytically written as CLSs.
We have shown in Sec.~\ref{sec:2nrootsl} that, for the zero flux case, one can construct the CLSs of all the flat bands of the $\sqrt[2^n]{\text{SL}}$, corresponding to the different harmonics of the linkage eigenstates, by directly generalizing the known form of the CLS of the Lieb lattice, as illustrated in Fig.~\ref{fig:6}(a).
The same reasoning can be applied for the case of a finite  magnetic flux per plaquette.
First, we start by noticing that the CLSs of the Lieb lattice have already been shown to survive the introduction of flux by extending over two plaquettes instead of one \cite{Lopes2011,Gouveia2016}.
For the gauge configuration defined in Fig.~\ref{fig:6}(b), these CLSs have the form shown at the left in Fig.~\ref{fig:6}(c).
Then, as before for the zero flux case, all CLSs of the $\sqrt[2^n]{\text{SL}}$ can be constructed by simply translating the phases of the finite components at the eight sites with equal weight in the green sublattice of the Lieb CLS into global phases for the different harmonics of the respective linkages of the $\sqrt[2^n]{\text{SL}}$, as exemplified by the left$\to$right passage in Fig.~\ref{fig:6}(c) where, again, the parity of the $j^{th}$ harmonic has to be taken into account, since odd (even) harmonics are (anti-)symmetric in relation to the inversion-axis of the linkage.

It is important to underline that the perpendicular CLS is also a solution for the cases shown in Fig.~\ref{fig:6}(c), that is, each CLS can be made to occupy instead the top left and bottom right plaquette only (notice that this perpendicular solution will have a different phase configuration).
Finally, Fig.~\ref{fig:6}(c) also highlights the trivial Chern topology of the flat bands, since varying the flux affects the phase configuration of the CLSs along the lattice but not their global cardinality, such that $C=0$ from (\ref{eq:cherndiscrete}) for each flat band.

\section{Experimental implementation in optical lattices}
\label{sec:exp_impl}

In this section, we describe different schemes to produce the lattice structures described previously in experimental implementations and subsequently probe their novel signatures.
Since it is easier to implement in an experiment, we assume throughout this section that the hopping terms are uniform and equal to $J$ across the linker and linkage sites of the different $2^n$-root lattices, but all the discussion below would also be applicable to the case with non-uniform couplings as in the sections above.

Optical potentials can be manipulated in different ways to obtain the decorated lattices studied in this article.
One possibility is to use digital micromirror devices \cite{Gauthier:16} or spatial light modulators \cite{Bergamini:04} to directly implement the desired structure. As we detail in Appendix \ref{app:exp_realization}, this usually involves introducing potential offsets in targeted sites of each unit cell of an existing square lattice, to effectively remove them from the lattice potential experienced by the atoms.

\begin{figure}[tb]
	\begin{centering}
		\includegraphics[width=0.48 \textwidth]{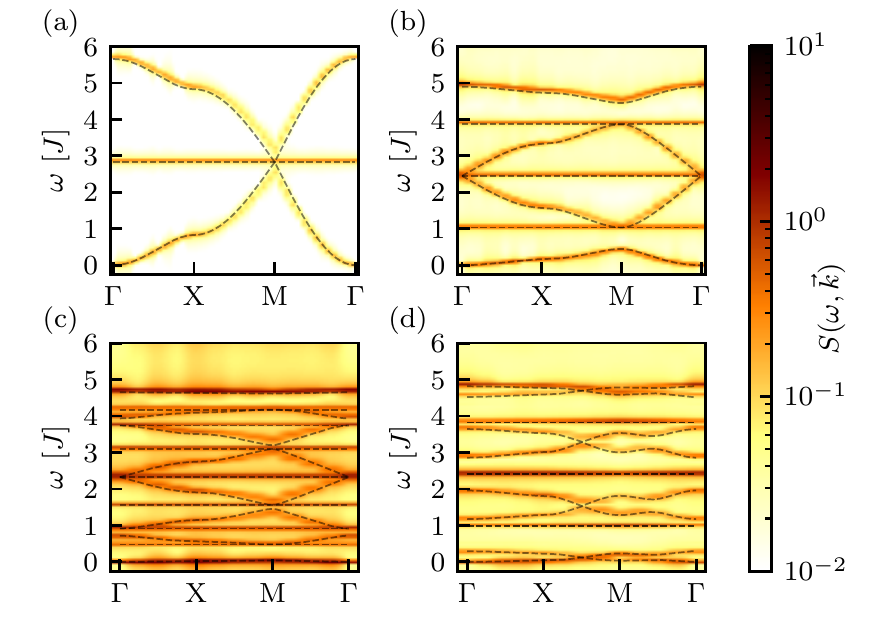}
		\par\end{centering}
	\caption{
		Dynamical structure factor $S(\omega, \vec{k})$ for finite lattices with different configurations in a logarithmic plot.
		The dashed gray lines are the energy spectrum obtained from exact diagonalization in an infinite periodic lattice.
		(a) $n=1$, $\phi=0$, $L_x=L_y=30$,
		(b) $n=2$, $\phi=0$, $L_x=L_y=20$,
		(c) $n=3$, $\phi=0$, $L_x=L_y=10$,
		(d) $n=2$, $\phi=\pi$, $L_x=L_y=10.$
	}
	\label{fig:11}
\end{figure}

Optical lattice implementations offer the possibility to detect the repeating structure of the energy bands for the $2^n$-root lattices by measuring the dynamical structure factor (DSF) via Bragg spectroscopy.
These types of measurements have been experimentally demonstrated for a BEC~\cite{PhysRevLett.83.2876} and for free fermions~\cite{PhysRevLett.128.100401}, where it is possible to extract the spectral information directly in frequency and momentum space.
In our numerical simulations we begin in the single-particle ground state and perform unitary time evolution.
We then compute time-separated density-density correlation functions $\langle n_{\vec{r}_1,t} n_{\vec{r}_2,0} \rangle$ and Fourier transform them in time and space to obtain the DSF,
\begin{equation} \label{equ_DSF}
    S(\omega, \vec{k}) = \sum_{t} e^{i \omega t} \sum_{\vec{r}_1, \vec{r}_2} e^{i \vec{k} (\vec{r}_1-\vec{r}_2)}
	\langle n_{\vec{r}_1,t} n_{\vec{r}_2,0} \rangle.
\end{equation}
For each value of $\vec{k}$, we expect $S(\omega, \vec{k})$ to have peaks at values of $\omega$ that correspond to energy differences between the ground state and one of the single-particle energy bands.
In Appendix \ref{app:dsf}, we provide more details of how we perform these calculations in finite lattices with unit cells of more than one site.
In Fig.~\ref{fig:11}, we plot the results of our simulations for different lattices with open boundary conditions, with the logarithmic color scale indicating the DSF values and the dashed lines signaling the theoretical values of the energy bands shifted by the ground state energy.
In particular, the plots in Fig.~\ref{fig:11}(a), (b) and (c) correspond respectively to the fluxless modified (due to the uniform couplings) $\sqrt[2^1]{\text{SL}}$, $\sqrt[2^2]{\text{SL}}$ and $\sqrt[2^3]{\text{SL}}$ lattices.
In all cases, we observe that the regions in the $\vec{k}$-$\omega$ space where the maxima of $S(\omega,\vec{k})$ occur agree very well with the shape of the different energy bands, whose periodic structure is clearly revealed.
In Fig.~\ref{fig:11} (d), we also compare for the $\sqrt[2^2]{\text{SL}}$ lattice to the case with flux $\phi=\pi$, which can be realized in these experimental architectures by employing Raman-assisted tunneling processes~\cite{PhysRevLett.107.255301}.
We again observe that the DSF can be used to reconstruct the repeating band structure with high accuracy.

Even though the plots in Fig.~\ref{fig:11} have been produced with 600 equally spaced values of $\omega$ and up to 50 values of the momentum $\vec{k}$, in an experiment it would suffice to sample only a few points in ($\vec{k},\omega$) space to detect the repeating nature of the band structure and the presence of multiple flat bands.


\begin{figure}[tb]
	\begin{centering}
		\includegraphics[width=0.48 \textwidth]{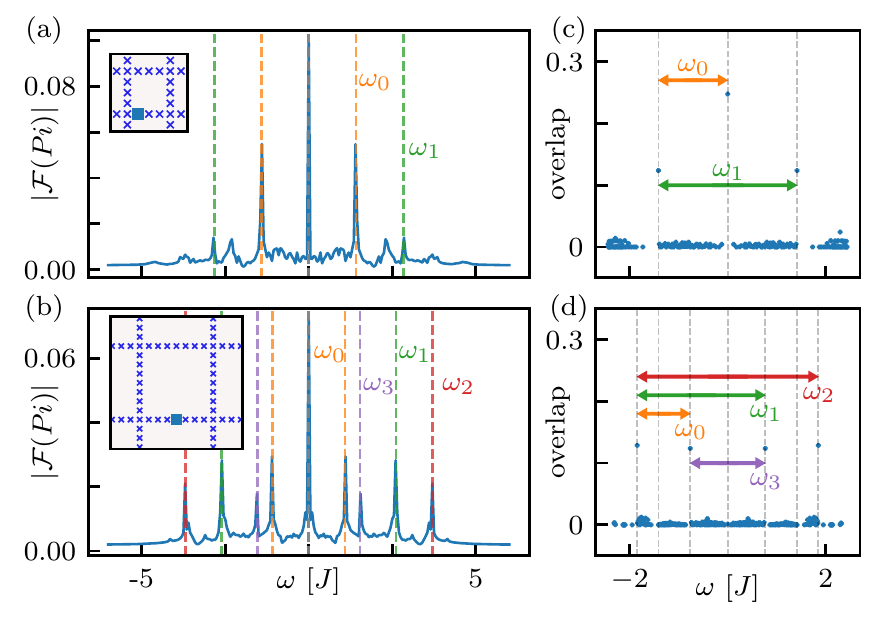}
		\par\end{centering}
	\caption{
		Fourier transform of the time evolution of the survival probability of a single atom prepared in a  linkage site of (a) a $\sqrt[2^2]{\text{SL}}$ and (b) a $\sqrt[2^3]{\text{SL}}$ lattice.
		In each case the positions of the atom at time $t=0$ inside a unit cell are marked in the insets as a blue square.
		The dashed colored lines mark the energy difference of two flat bands corresponding to the colored arrows in plots (c) and (d), which show the overlaps $|d_j|^2$ of the initial state shown respectively in (a) and (b) with the eigenstates of each lattice as a function of their energy.
		The largest overlaps correspond to CLSs localized at the AB cage where the particle is prepared.
	}
	\label{fig:12}
\end{figure}

Another possibility to probe the multiple flat band structure is to prepare single particles in one of the linkage sites and observe the time evolution of the probability that the particle remains in that site.
The initial state wave function at site $i$ can be written as a superposition of the single-particle eigenstates $\ket{\phi_j}$,
\begin{equation}
\ket{\Psi_i}=\sum_j d_j \ket{\phi_j},\ \ \sum_j |d_j|^2=1.
\end{equation}
Thus, the probability that the particle is found at the initial site after a time $t$ can be expressed as
\begin{equation}
P_i(t)=\left|\langle\Psi_i|\Psi_i(t)\rangle\right|^2=\left|\sum_j|d_j|^2e^{-i E_j t}\right|^2.
\label{timeevsinglepart}
\end{equation}
Since $\ket{\Psi_i}$ is completely localized in a linkage site, its overlaps $|d_j|^2$ with the flat band CLSs are very large compared to the overlaps with the dispersive states.
Thus, after preparing a single particle in a linkage site a macroscopic fraction of the wave function remains in the AB cage defined by the CLSs with which it overlaps, and the probability $P_i(t)$ of finding the particle in the initial site is dominated by terms which oscillate at angular frequencies $\omega_k=\left|E_i-E_j\right|$ that correspond to the energy differences between the flat bands.
In order to illustrate this effect, in Fig.~\ref{fig:12}(a) we plot the Fourier transform of $P_i(t)$ for an initial state prepared in a linkage site of the bulk of a $\sqrt[2^2]{\text{SL}}$ lattice, as shown in the left inset.
As discussed previously, this lattice has three flat bands of energies $E=\pm \sqrt{2}J,0$ (note that we have omitted the $\sqrt{2}$ factor in the linkage hopping terms).
As shown in Fig.~\ref{fig:12}(c), where we plot the overlap of the localized state shown in the inset of Fig.~\ref{fig:12}(a) with the different eigenstates as a function of their energies, the initial state has much larger components on the flat band CLSs than on the rest of the states.
Therefore, the dominant (non-zero) oscillation frequencies of $P_i(t)$ are $\omega_0=(\sqrt{2}-0)J=\sqrt{2}J$ and $\omega_1=(\sqrt{2}-(-\sqrt{2}))J=2\sqrt{2}J$, as can be seen in Fig.~\ref{fig:12}(a).

In Fig.~\ref{fig:12}(b), we show the Fourier transform of $P_i(t)$ for an initial state prepared in the central site of a linkage chain in the bulk of a $\sqrt[2^3]{\text{SL}}$ lattice, as shown in the left inset.
In Fig.~\ref{fig:12}(d), we show the overlaps of this initial state with all the eigenstates of the lattice as a function of their energy.
Even though this system has seven flat band states as discussed in the previous sections, because the initial state is prepared at the center of the linkage chain it only overlaps with four of the flat band CLSs, since the three even $j$ harmonics in (\ref{eq:bulkflateigssl}) have nodes at the central site of the linkage.
Since these overlaps are again much larger than those with the rest of the eigenstates, the Fourier spectrum of the survival probability is dominated by the four possible combinations of frequency differences between the flat bands $\omega_0,\omega_1,\omega_2,\omega_3$, as shown in Fig.~\ref{fig:12}(d).
\\
This type of real-space measurement of survival probabilities could be performed in different experimental platforms.
One possibility would be to use Rydberg atom arrays, which allow for the simulation of single-particle dynamics in a wide range of lattice geometries and offer single-site resolution of the excitations \cite{lienhard2020realization,de2019observation,weber2018topologically,wu2022synthesizing}.
Alternatively, one could use a quantum gas microscope to image individual atoms in an optical lattice \cite{Quan_Gas_mic,bakr2009quantum}.
In either case, to obtain frequency spectra with sufficient resolution to identify the peaks corresponding to the flat bands it would be necessary to evaluate the survival probability at $\sim 100$ different times, which would require $\sim 100$ measurements per probability evaluation.

\section{Conclusions}
\label{sec:conclusions}

We have studied the Hofstadter regime of $2^n$-root topological insulators.
We started by considering a quasi-one-dimensional model, whose square-root version is the diamond chain, and showed that, as $n$ is increased, several flux insensitive flat bands appear in the Hofstadter spectrum.
The relation between the spectrum of a $2^n$-root chain and its lower root-degree versions, obtained by a scheme centered on successive squaring operations on the starting Hamiltonian, was derived and analyzed.

Then, we proceeded to study the $2^n$-root versions of the two-dimensional square lattice, whose square-root version was shown to correspond to the Lieb lattice.
Analogously to the case of the lower dimensional chain, the Hofstadter spectrum revealed here that, for $n>1$, a subspace composed of multiple flat bands at different energy levels was seen to survive the introduction of an arbitrary flux per plaquette.
The compact localized states of these flat bands respond to the introduction of flux by spreading from one to two plaquettes, diagonally connected by a shared corner site with a node in the wave function, which constitutes a common Aharonov-Bohm cage within which each of these states corresponds to a different harmonic, with regard to its decoupled linkage chains.

The Hofstadter diagram of the $2^n$-root square lattice was shown to display $2^n$ vertically placed Hofstadter butterflies, with each pair separated by a flux insensitive flat band, for a total of $2^n-1$.
Each of the butterflies in the kaleidoscope has been found to be a topologically equivalent replica of the original Hofstadter butterfly, in what concerns the Chern number at each of the energy gaps.
Therefore, the $n$ index identifying the root-degree of the lattice can be understood as another fractal dimension, where the Hofstadter spectrum is replicated along the vertical energy direction with increasing $n$, as shown in Fig.~\ref{fig:7}, in a way that resembles a film roll with an in-depth bent where each ``frame'', delimited by successive flat bands, is a copy of the same butterfly.

We also have seen that the repeating structure of the spectrum can, in principle, be resolved in experimental realizations of the $2^n$-root square lattice, such as ultracold atoms in optical lattices, by measuring the dynamical structure factors.
Furthermore, the flat bands can be studied in more detail in experimental systems that allow for site resolved measurements, e.g., neutral atoms in optical tweezers.
This can be performed by time evolving single excitations that have large overlaps with the flat band states.
By looking at the oscillation of the single particle one can then infer the relative energies of the different flat band states.

As for future work, our $2^n$-root square lattice can be readily generalized to a hypercubic lattice of dimension $d>2$, by substituting the $\sqrt{2}$ factor appearing in the hopping terms of the linkage (see Fig.~\ref{fig:4}) with $\sqrt{d}$.
However, it is not yet clear if an analytical expression for the eigenstates of the flat bands, based on the generalization of the intertwiner relation in (\ref{eq:intertwinerprime}), is available for these higher-dimensional systems, and therefore remains an open question.
At the same time, for $d>2$ there is an additional degree of freedom for producing the Hofstadter spectrum, related to the choice of direction along which the magnetic flux pierces the lattice.
This may lead to new phenomena, such as direction dependent Aharonov-Bohm caging for the flat band subspace.
We note that recent developments that can be related to this line of inquiry have already been made in certain types of three-dimensional decorated Lieb lattices \cite{Liu2020,Liu2021,Liu2022}.
Our work can also be developed in another direction, namely that of generalizing our results in $2^n$-root topological lattices to simply $n$-root systems, for any $n\in\mathbb{N}$.
This requires, as we will show elsewhere, the introduction of non-Hermiticity in the models in a very specific way, and therefore falls outside the scope of the present study.

\section*{Acknowledgments}
\label{sec:acknowledments}

This work was developed within the scope of the Portuguese Institute for Nanostructures, Nanomodelling and Nanofabrication (i3N) projects No.~UIDB/50025/2020, No.~UIDP/50025/2020 and No.~LA/P/0037/2020, and funded by FCT-Portuguese Foundation for Science and Technology through the project PTDC/FIS-MAC/29291/2017.
Work at the University of Strathclyde was supported by the EPSRC Programme Grant DesOEQ (EP/P009565/1).
AMM acknowledges financial support from the FCT through the work contract No.~CDL-CTTRI-147-ARH/2018 and from
i3N through the work Contract No.~CDL-CTTRI-46-SGRH/2022.


\appendix

\section{Off diagonal Hamiltonian blocks}
\label{app:hblocks}
Here, we give the explicit matrix expression for the off-diagonal block of the quasi-1D Hamiltonian in (\ref{eq:hamilcl2ton}),
\begin{widetext}
\begin{equation}
	h_{\sqrt[2^n]{\text{CL}}}(k)=
	\begin{pmatrix}
		1&&&&e^{-ik}&1&&&&e^{-i(k+\phi)}
		\\
		\sqrt{2}&\sqrt{2}&
		\\
		&\sqrt{2}&\sqrt{2}&
		\\
		&&\ddots&\ddots
		\\
		&&&\sqrt{2}&\sqrt{2}&&
		\\
		&&&&&\sqrt{2}&\sqrt{2}&
		\\
		&&&&&&\sqrt{2}&\sqrt{2}
		\\
		&&&&&&&\ddots&\ddots
		\\
		&&&&&&&&\sqrt{2}&\sqrt{2}
	\end{pmatrix},\ \ n\geq 1,
\end{equation}
\end{widetext}
and of the 2D Hamiltonian in (\ref{eq:hamilsl2ton}),
\begin{widetext}
\begin{equation}
h_{\sqrt[2^n]{\text{SL}}}(\mathbf{k})=
\begin{pmatrix}
	1&&&&e^{-ik_y}&1&&&&e^{-ik_x}
	\\
	\sqrt{2}&\sqrt{2}&
	\\
	&\sqrt{2}&\sqrt{2}&
	\\
	&&\ddots&\ddots
	\\
	&&&\sqrt{2}&\sqrt{2}&&
	\\
	&&&&&\sqrt{2}&\sqrt{2}&
	\\
	&&&&&&\sqrt{2}&\sqrt{2}
	\\
	&&&&&&&\ddots&\ddots
	\\
	&&&&&&&&\sqrt{2}&\sqrt{2}
\end{pmatrix},\ \ n\geq 1.
\end{equation}
\end{widetext}
In both cases, all entries not shown are zeros.

\section{Intertwiner of the $\sqrt[2^n]{\text{SL}}$ with flux}
\label{app:intertwinerslflux}
Considering the $\sqrt[2^n]{\text{SL}}$ with a rational reduced flux per plaquette given by $\phi=2\pi\frac{p}{q}$, with $p$ and $q$ assumed to be coprime integers, the magnetic unit cell has the form of Fig.~\ref{fig:10}, which is $q$ times the size of the structural unit cell in Fig.~\ref{fig:4}.
For this choice of unit cell and gauge configuration, the bulk Hamiltonian of the model can be written in the generalized $\mathcal{B}_A$ basis, with the site ordering following (\ref{eq:basisa}) within each of the $q$ structural unit cells and increasing from left to right, as a block tridiagonal matrix with nontrivial anti-diagonal corner blocks enforcing PBC along the $x$ direction,
\begin{widetext}
		\begin{eqnarray}
	H^{(p,q)}_{\sqrt[2^n]{\text{SL}},A}(\mathbf{k})&=&
	\begin{pmatrix}
    H^{(p,q,0)}_{\sqrt[2^n]{\text{SL}},A}(k_y)&R&&S(k_x)
    \\
	R^T&H^{(p,q,1)}_{\sqrt[2^n]{\text{SL}},A}(k_y)&\ddots
	\\
	&\ddots&\ddots&
	\\
	&&H^{(p,q,q-2)}_{\sqrt[2^n]{\text{SL}},A}(k_y)&R
	\\
	S^\dagger(k_x)&&R^T&H^{(p,q,q-1)}_{\sqrt[2^n]{\text{SL}},A}(k_y)
	\end{pmatrix},
	\\
	H^{(p,q,l)}_{\sqrt[2^n]{\text{SL}},A}(k_y)&=&
	\begin{pmatrix}
	0&\mathbf{u}_{2^n}(k_y,l)&\mathbf{v}_{2^n}
	\\
	\mathbf{u}^{\dagger}_{2^n}(k_y,l)&H^{\text{SL}}_{\text{Linkage},2^n}&O_{2^{n}-1}
	\\
	\mathbf{v}^{\dagger}_{2^n}&O_{2^{n}-1}&H^{\text{SL}}_{\text{Linkage},2^n}&
	\end{pmatrix},
	\\
	R&=&
	\underbrace{		\begin{pmatrix}
		0&0&\dots&0&0
		\\
		0&&&&0
		\\
		\vdots&&\ddots&&\vdots
		\\
		0&&&&0
		\\
		1&0&\dots&0&0
		\end{pmatrix}}_{2^{n+1}-1},
		\ \ \ \ \ \ \ S(k_x)=
	\underbrace{		\begin{pmatrix}
			0&0&\dots&0&e^{-ik_x}
			\\
			0&&&&0
			\\
			\vdots&&\ddots&&\vdots
			\\
			0&&&&0
			\\
			0&0&\dots&0&0
	\end{pmatrix}}_{2^{n+1}-1},
	\end{eqnarray}
\end{widetext}
where $\mathbf{u}_{2^n}(k_y,l)=-\sqrt[2^n]{J}(e^{i2\pi l\frac{p}{q}},0,\dots,0,e^{-ik_y})$ and $\mathbf{v}_{2^n}=-\sqrt[2^n]{J}(1,0,\dots,0,0)$ are vectors of size $2^{n}-1$, $H^{\text{SL}}_{\text{Linkage},2^n}$ is defined in (\ref{eq:hamilt2nsl}) and all entries not shown are zeros.
$H^{(p,q)}_{\sqrt[2^n]{\text{SL}},A}(\mathbf{k})$ is a $q(2^{n+1}-1)\times q(2^{n+1}-1)$ matrix, such that its diagonalization yields $q(2^{n+1}-1)$ energy bands.

Our goal is to find an intertwiner that satisfies
\begin{equation}
H^{(p,q)}_{\sqrt[2^n]{\text{SL}},A}(\mathbf{k})G_{2^n}^{(p,q)}(\mathbf{k})=G_{2^n}^{(p,q)}(\mathbf{k})H^{\text{SL}}_{\text{Linkage},2^n}.
\label{eq:intertwinerprimeflux}
\end{equation}
We assume the intertwiner generalizes (\ref{eq:intertwinersl}) as
\begin{equation}
G_{2^n}^{(p,q)}(\mathbf{k})=\begin{pmatrix}
\vec{O}_{2^{n}-1}
\\
\alpha_0 I_{2^{n}-1}+\gamma_0 J_{2^{n}-1}
\\
\beta_0 I_{2^{n}-1}+\delta_0 J_{2^{n}-1}
\\
\vec{O}_{2^{n}-1}
\\
\alpha_1 I_{2^{n}-1}+\gamma_1 J_{2^{n}-1}
\\
\beta_1 I_{2^{n}-1}+\delta_1 J_{2^{n}-1}
\\
\vdots
\\
\vec{O}_{2^{n}-1}
\\
\alpha_{q-1} I_{2^{n}-1}+\gamma_{q-1} J_{2^{n}-1}
\\
\beta_{q-1} I_{2^{n}-1}+\delta_{q-1} J_{2^{n}-1}
\end{pmatrix},
\label{eq:intertwinerslflux}
\end{equation}
which is a rectangular matrix of size $q(2^{n+1}-1)\times(2^n-1)$ and where the zero rows impose nodes at the blue sites in Fig.~\ref{fig:10} with coordination number of four.
Inserting (\ref{eq:intertwinerslflux}) back in (\ref{eq:intertwinerprimeflux}) leads to a system of equations,
\begin{equation}
	\begin{cases}
		\alpha_0+e^{-ik_y}\gamma_0+\beta_0+e^{-ik_x}\delta_{q-1}=0
		\\
		\gamma_0+e^{-ik_y}\alpha_0+\delta_0+e^{-ik_x}\beta_{q-1}=0
	\end{cases},
\label{eq:recursivea}
\end{equation}
and
\begin{equation}
	\begin{cases}
		e^{i2\pi l\frac{p}{q}}\alpha_l+e^{-ik_y}\gamma_l+\beta_l+\delta_{l-1}=0
		\\
		e^{i2\pi l\frac{p}{q}}\gamma_l+e^{-ik_y}\alpha_l+\delta_l+\beta_{l-1}=0
	\end{cases},
    \ l=1,2,\dots,q-1,
    \label{eq:recursiveb}
\end{equation}
where (\ref{eq:recursivea}) and (\ref{eq:recursiveb}) are recursive due to the last term on the left-hand side of each equation.
Upon setting all $\beta_i=\delta_i:=1$, the other coefficients can be determined to read as
\begin{equation}
	\begin{cases}
		\beta_i=\delta_i:=1,\ \  \ \ i=0,1,\dots,q-1,
		\\
		\alpha_0=\gamma_0=-\frac{1+e^{-ik_x}}{1+e^{-ik_y}},
		\\
		\alpha_l=\gamma_l=-\frac{2}{e^{i2\pi l\frac{p}{q}}+e^{-ik_y}}, \ \ l=1,2,\dots,q-1,
	\end{cases}
\end{equation}
where $k_y\neq\pi\vee\pi (1-2l\frac{p}{q}) \mod 2\pi$ is assumed, and different choices for $\beta_i$ and $\delta_i$ have to be made at these points to avoid singularities.
Finally, the $j^\text{th}$ eigenstate within the flat band subspace of the bulk Hamiltonian $H^{(p,q)}_{\sqrt[2^n]{\text{SL}},A}(\mathbf{k})$ can be written as
\begin{equation}
	\psi^{(p,q)}_{2^n,j}(\mathbf{k})=\frac{1}{\mathcal{N}_{2^n}^{j,(p,q)}(\mathbf{k})}G_{2^n}^{(p,q)}(\mathbf{k})\varphi^\prime_{2^n,j},
	\label{eq:bulkflateigssl2}
\end{equation}
with $\mathcal{N}_{2^n}^{j,(p,q)}(\mathbf{k})$ the normalization constant and $\varphi^\prime_{2^n,j}$ defined in (\ref{eq:schroedlinkagesl}).

\section{Possible experimental realization}
\label{app:exp_realization}

In this appendix, we want to discuss one possibility to implement a $2^n$-root lattice in optical potentials.
For simplicity, we will only consider the $n=2$ case.

\begin{figure}[ht]
	\begin{centering}
		\includegraphics[width=0.48 \textwidth]{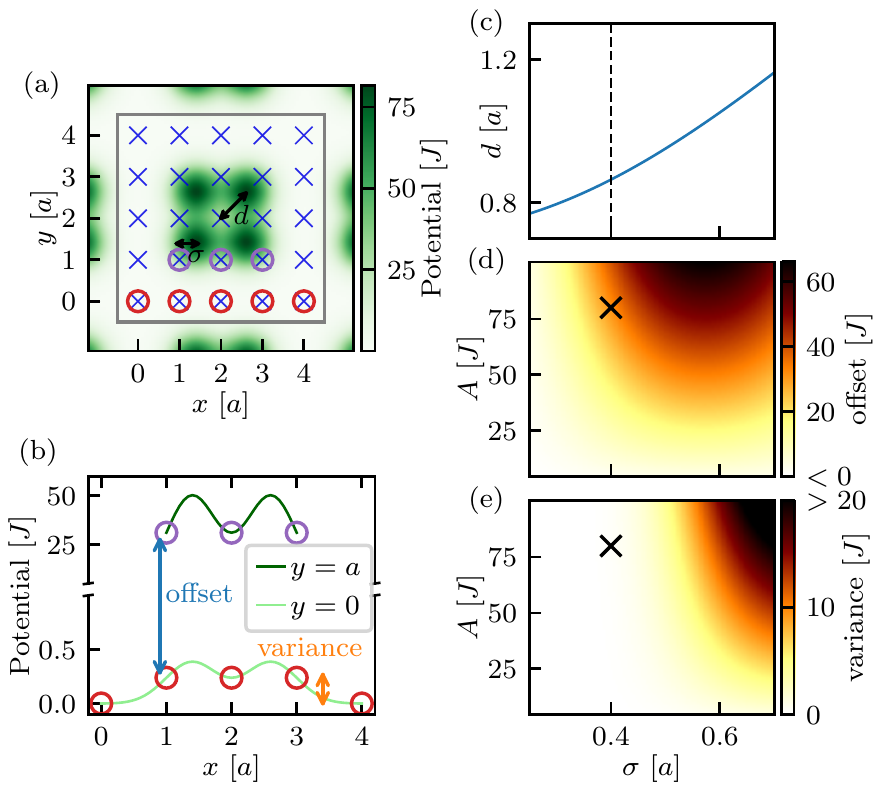}
		\par\end{centering}
	\caption{
		(a) Additional optical potential at one unit cell to remove the inner sites and create an effective $\sqrt[2^2]{\text{SL}}$ lattice.
		(b) The potential offset for the sites belonging to the $\sqrt[2^2]{\text{SL}}$ lattice ($y=0$) and the adiabatically removed sites ($y=a$).
		(c) Optimized displacement $d$ as function of the beam width $\sigma$ to minimize the variance of the three linkage sites to $0$.
		(d) The potential offset between the linkage sites and the adiabatically removed sites and
		(e) the potential variance between the linkage sites and the corner sites as function of the beam width $\sigma$ and the light intensity $A$.
		The black cross denotes an example point to minimize the variance while maximizing the offset.
	}
	\label{fig:13}
\end{figure}

We start with a square optical lattice with lattice spacing $a$.
To obtain the $\sqrt[2^2]{\text{SL}}$ lattice we have to adiabatically remove the nine inner sites for each unit cell [see Fig.~\ref{fig:13}(a)].
This can be achieved by applying additional beams (e.g., via a spatial light modulator) onto the lattice to obtain a potential offset at these sites.
Due to symmetry considerations, it is best to apply four beams for each unit cell, which are symmetrically displaced from the center with the distance $d$ and have a Gaussian intensity profile
\begin{equation}
    I(r)=Ae^{\frac{-r^2}{2\sigma^2}}.
\end{equation}

One can apply the beams in such a way that all inner sites get an as large as possible energy offset, while the sites belonging to the $\sqrt[2^2]{\text{SL}}$ lattice are only shifted minimally, see Fig.~\ref{fig:13}(b).

In fact, for all beam widths of interest ($\sigma \approx 0.5 a$) one can find a displacement $d$, such that the sites at $y=0$ and $x/a = 1,2,3$, marked by red circles in Fig.~\ref{fig:13}(a) and \ref{fig:13}(b), have the exact same energy offset.
This optimized displacement $d$ as a function of the beam width $\sigma$ can be seen in Fig.~\ref{fig:13}(c).
Therefore, the flat band states, which only live on these linkages, do not get disturbed by such an additional optical potential.

Furthermore, we now look at the energy offset of the inner sites compared to the linkage sites [Fig.~\ref{fig:13}(d)] as well as the variance of the energy offset on the sites of the $\sqrt[2^2]{\text{SL}}$ lattice [Fig.~\ref{fig:13}(e)].
For simplicity, we define the variance as maximal difference of the optical potential at the corner site to the linkage sites.
To get the best effective $\sqrt[2^2]{\text{SL}}$ lattice we want to maximize the offset while the variance is as small as possible.
This can be done for example by choosing $\sigma = 0.4 a$ and $A = 80 J$, which is marked by a black cross in Fig.~\ref{fig:13}(d) and (e).
We checked for these parameters, as well as for a small displacement of the grid ($\sim 0.1 a$), that the energy bands of the system only get disturbed slightly and all results of Sec.~\ref{sec:exp_impl} are still observable.

In order to go to $n>2$, the same procedure followed in this section can be applied, only now more inner sites have to be removed by the beams.
E.g., for $n=3$, an inner square of $7\times 7$ sites has to be removed.
One possibility to do so is by applying $6\times 6$ beams per unit cell.

\section{Details on the dynamical structure factor}
\label{app:dsf}

The dynamical structure factor $S(\omega, \vec{k})$ in (\ref{equ_DSF}) can be obtained by calculating the time and space Fourier transform of the density-density correlation functions $\langle n_{\vec{r}_1,t} n_{\vec{r}_2,0} \rangle$.
However, this does not use any information about the lattice and its unit cells.
For the square lattice (power $n=0$) a unit cell consists only of one site and therefore the distance between two unit cells is the same as between two sites: $a$.
Therefore, the momentum space for such a lattice is defined as $-\pi/a \leq k_i < +\pi/a$.


If we now consider the $\sqrt{\text{SL}}$ lattice, its unit cell consists of two sites in $x$ (and also $y$) direction.
Therefore, the shift between two unit cells is now $2 a$ and in the case of an infinite periodic lattice the eigenstates will be labelled by their momentum ranging from $-\pi/2a$ to $+\pi/2a$.

Nevertheless, we still can resolve the atom by each individual site (and not just the unit cell).
Naturally the dynamical structure factor calculation will still obtain momentum values in the range of $-\pi/a$ to $+\pi/a$.
But all momentum states with $\abs{k} > \pi/2a$ correspond to wavelengths inside a single unit cell.
To map these results of the dynamical structure factor to the results of an infinite periodic lattice we can fold the range from $+\pi/2a$ to $+\pi/a$ back to the range $+\pi/2a$ to $0$.

\begin{figure}[ht]
	\begin{centering}
		\includegraphics[width=0.48 \textwidth]{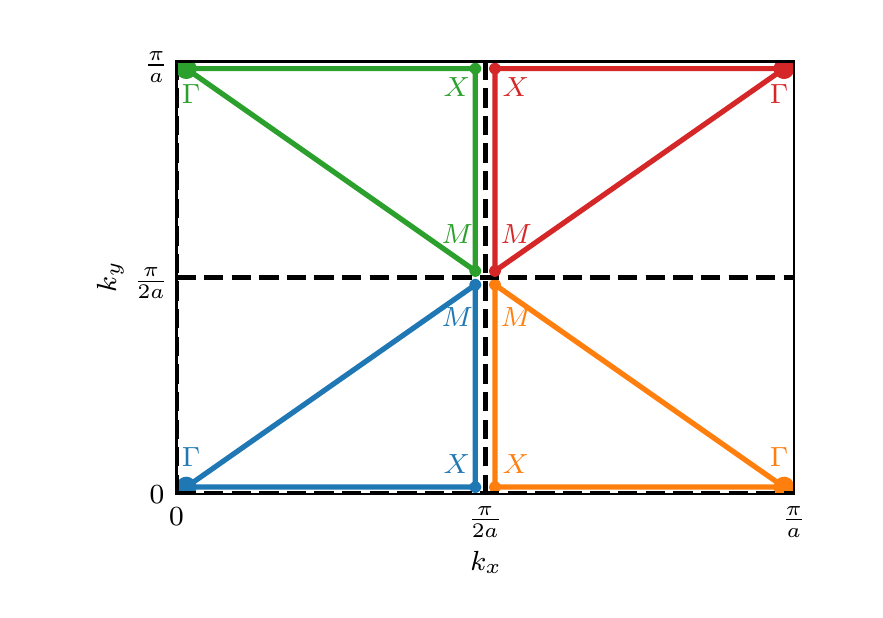}
		\par\end{centering}
	\caption{
		Sketch of the folding process of the Brillouin zone for $n=1$.
		Since our unit cell has the size $2 a$, we are interested in the effective Brillouin zone up to momenta $k_i = \frac{\pi}{2a}$.
		However, to avoid losing any information of the points outside this effective Brillouin zone we fold the outer parts into the effective Brillouin zone and thus identify the marked points.
		To calculate the dynamical structure factor at one point in the effective Brillouin zone we therefore sum up all the dynamical structure factors that identify with a given $k$-point.
	}
	\label{fig:14}
\end{figure}

For the exemplary high-symmetry ($\Gamma - X - M - \Gamma$) path this is shown in Fig.~\ref{fig:14}.
For the result of the dynamical structure factor on the high-symmetry path all the results of the four mirrored paths inside the big Brillouin zone are added together.
This scheme can also be generalized for higher square-root lattices ($n\geq1$) and is implicitly done in Fig.~\ref{fig:11}.
Therefore, for each data point in Fig.~\ref{fig:11}, $4^n$ dynamical structure factors are calculated and summed together.

\newpage


\bibliography{hofst}

\end{document}